\documentclass{aastex}

\shorttitle{Evolution of broad line regions} \shortauthors{Zhu,
Zhang \& Tang}

\begin{document}

\title{Evidence for an intermediate line region in AGN's inner torus region and
its evolution from narrow to broad line Seyfert I galaxies}

\author{Ling Zhu\altaffilmark{1}, Shuang-Nan Zhang\altaffilmark{1,2,3}, Su-Min Tang\altaffilmark{4}}

\altaffiltext{1}{Department of Physics and Tsinghua Center for
Astrophysics, Tsinghua University, Beijing 100084, China;
zhul04@mails.tsinghua.edu.cn, zhangsn@tsinghua.edu.cn}
\altaffiltext{2}{Key Laboratory of Particle Astrophysics, Institute
of High Energy Physics, Chinese Academy of Sciences, P.O. Box 918-3,
Beijing 100049, China} \altaffiltext{3}{Physics Department,
University of Alabama in Huntsville, Huntsville, AL 35899,
USA}\altaffiltext{4}{Harvard-Smithsonian Center for Astrophysics, 60
Garden St, Cambridge, MA 02138, USA}

\begin{abstract}
A two-components model for Broad Line Region (BLR) of Active
Galactic Nuclei (AGN) has been suggested for many years but not
widely accepted (e.g., Hu et al. 2008; Sulentic et al. 2000;
Brotherton et al. 1996; Mason et al. 1996). This model indicates
that the broad line can be described with superposition of two
Gaussian components (Very Broad Gaussian Component (VBGC) and
InterMediate Gaussian Component (IMGC)) which are from two
physically distinct regions; i.e., Very Broad Line Region (VBLR) and
InterMediate Line Region (IMLR). We select a SDSS sample to further
confirm this model and give detailed analysis to the geometry,
density and evolution of these two regions. Micro-lensing result of
BLR in J1131-1231 and some unexplained phenomena in Reverberation
Mapping (RM) experiment provide supportive evidence for this model.
Our results indicate that the radius obtained from the emission line
RM normally corresponds to the radius of the VBLR, and the existence
of the IMGC may affect the measurement of the black hole masses in
AGNs. The deviation of NLS1s from the M-sigma relation and the Type
II AGN fraction as a function of luminosity can be explained in this
model in a coherent way. The evolution of the two emission regions
may be related to the evolutionary stages of the broad line regions
of AGNs from NLS1s to BLS1s. Based on the results presented here, a
unified picture of hierarchical evolution of black hole, dust torus
and galaxy is proposed.
\end{abstract}


\section{Introduction}

Type I AGNs are often classified into two subclasses according to
the Full Width at Half Maximum (FWHM) of their broad H$\beta$ lines.
Those AGNs with FWHM greater than approximately 2000
km$\cdot$s$^{-1}$ are called Broad Line Seyfert 1 (BLS1), and those
with FWHM less than approximately 2000 km$\cdot$s$^{-1}$ are called
Narrow Line Seyfert 1 (NLS1). There is a tight correlation between
the black hole mass and the stellar velocity dispersion of the bulge
of normal galaxies, the so called M-sigma relation (e.g. Magorrian
et al. 1998; Gebhardt,  et al. 2000;  Tremaine et al. 2002; Marconi
\& Hunt 2003; Ferrarese \& Ford 2005). BLS1s are found also to
follow this relation well (Greene \& Ho 2006). However, NLS1s seem
to deviate from this relation; they seem to have much smaller masses
or much higher stellar velocity dispersions (e.g., Wang \& Lu 2001;
Bian \& Zhao 2004; Zhou et al. 2006; Komossa \& Xu 2007). {\it We
call this under-massive black hole problem of NLS1s.} Since there
are a lot of uncertainties on measuring the black hole mass and
sigma (Decarli et al. 2008; Komossa \& Xu 2007; Komossa 2008;
Komossa et al. 2008), it is difficult to determine whether the
deviation reflects an intrinsic mass or/and sigma difference between
the NLS1s and BLS1s, or it is only caused by the measurements. NLS1s
also show softer X-ray spectra, strong Fe II emission, rapid
continuum variation, and high accretion rate; the last one may
partly be caused by the incorrect measurement of the black hole mass
(e.g., Grupe \& Mathur 2004; Komossa 2008). However, Williams et al.
(2004) found some optically defined NLS1s are different from those
defined by X-ray selected NLS1s, suggesting that strong, ultrasoft
X-ray emission is not a universal characteristic of NLS1s and
challenging the current paradigm that NLS1s accrete near the
Eddington limit.

In the simple unified model for AGN, Type I and Type II objects
differ only in terms of the angle between the observer's line of
sight and the normal axis of a dusty torus (Urry \& Padovani 1995).
SEDs of QSOs also imply that the location of the inner wall of torus
is determined by the sublimation of dust by the central radiation
(e.g. Kobayashi et al. 1993; Elvis et al. 1994). It is widely
accepted that $R_{\rm{ torus}}\propto L^{0.5}$ (e.g., Elitzur 2006;
Peterson 2007; Nenkova 2008a), consistent with that predicted by the
toy model of Krolik \& Begelman (1988). Reverberation mapping (RM)
based on infrared emission also supports this relation (Suganuma et
al. 2006).

Many observations show that the fraction of type I AGNs increases
with luminosity (Ueda et al. 2003; Hasinger 2004, 2005; Steffen et
al. 2003, 2004; Barger et al. 2005; Wang et al. 2005), i.e.,
brighter AGNs are mostly type I AGNs. This is also equivalent
observationally to the situation that it is less probable to find
very luminous type II AGNs (mostly Seyfert II galaxies) in a random
sample of AGNs. {\it We call this under-populated luminous type II
problem of Seyfert galaxies.} However, the standard receding torus
model with constant height first suggested by Lawrence (1991) failed
to describe this luminosity fraction, and thus a modified receding
torus whose height slightly increases with luminosity with the
relation $h\propto L^{0.23}$ is needed (Simpson 2005). Based on the
luminosity fraction for Type I AGN, Wang et al. (2005) obtained a
relation between covering factor of the torus and luminosity,
$\log{C}=-0.17\log{L_{x}}-0.36$. The variation of covering factor
with luminosity may be totally caused by the receding of the inner
torus (Wang et al. 2005).

Clearly both the broad line regions and the dust tori of AGNs show strong dependence of
luminosity, yet the physical connections between these two key ingredients of AGN's
unification scheme are not yet fully understood. It is also not clear if the luminosity
dependence of dust torus and broad line region is simultaneously related to both
problems of under-massive black holes in NLS1 galaxies and under-populated luminous type II Seyfert galaxies.
Progress towards solving these problems may shed new lights to the
understanding the hierarchical evolution of AGN and its host galaxy, as well as the
feeding, fueling and growth of supermassive black holes. In this paper we attempt to
address both problems in a coherent way by studying a sample of SDSS AGNs with strong
broad emission lines.

The geometry and kinematics of broad line region in AGN have been
studied for about three decades but far from been fully understood.
There are mainly two kinds of views to interpret the structure of
BLR based on the profiles of balmer lines. One interprets the
profile as a Gaussian/Lorentz profile and BLR as a extended region
has different projected velocity distributions (e.g., Goncalves et
al. 1999; Collin et al. 2006). The other model is that the profile
is a superposition of double Gaussian profiles (Intermediate
Component+Very Broad Component in Hu et al. 2008; Broad
Component+Very Broad Component in Sulentic et al, 2000 and Marziani
et al. 2009. These two components will be re-named as InterMediate
Gaussian Component (IMGC)+Very Broad Gaussian Component(VBGC) in
this paper.), which come from two physically distinct emission
regions due to their substantial differences in the line widths
(e.g., Hu et al. 2008; Sulentic et al. 2000; Brotherton et al. 1996;
Mason et al. 1996); we call these two regions Very Broad Line Region
(VBLR) and InterMediate Line Region (IMLR). For the two-components
model, the results from these papers are not consistent with each
other and the physical interpretations to these two emission regions
are also different. Sulentic et al (2000) suggested that VBGC is
likely to arise in an optically thin region close to the central
source which is slightly red-shifted, whereas Hu et al. (2008)
concluded that IMGC is systematically red-shifted and may come from
the inflow. In addition, the study of partially obscured quasars
suggests that there is an inner narrow line region covered by dust
which may be consistent with our IMLR (Zhang et al. 2009). The
existence of these two emission regions need to be confirmed further
and their dynamical and physical properties also need to be further
studied. For this purpose, we select a SDSS sample with strong
H$\alpha$, H$\beta$ and H$\gamma$ lines, and decompose these balmer
lines based on the two-components model. Our goal is to provide
further evidence for this model, carry out more detailed analysis
about the dynamics and evolution of these two emission regions, and
ultimately to understand the above two problems (The under-massive
black hole problem of NLS1s and The under-populated luminous type II
problem of AGNs) in a coherent way.

 This paper is organized in the following way. Detailed
decomposition and FWHM measurement are described in Section 2. In
Section 3 we present H$\alpha$, H$\beta$, and H$\gamma$
decomposition and statistical analysis. In section 4 we confirm the
two-components model based on the line decomposition and give
detailed analysis of these two emission regions. Section 5 presents
other supporting evidence for this model. Section 6 shows that this
model provides a coherent interpretation to the two problems.
Section 7 is conclusion and discussion. The continuum luminosity and
some other parameters of AGNs are directly taken from Table 2 in La
Mura et al. (2007). FWHM is defined as FWHM of the whole line, FWHMi
means FWHM of the IMGC and FWHMb represents FWHM of the VBGC when
they are not clear in the context. The unit is km s$^{-1}$ for the
``width" of all lines and components, as well as the inferred
velocity, throughout this manuscript.

\section{Line decomposition and FWHM measurement}

The sample contains 90 objects with clear H$\alpha$, H$\beta$, and
H$\gamma$ line profiles that can be decomposed. They are selected by
La Mura et al. (2007) from SDSS 3 based on their balmer line intensities.
Narrow emission lines have already been removed by La Mura et al. (2007) from the spectra we
adopted, with templates extracted from OIII and with compatible width;
the detailed description on how they have produced the broad line spectral catalog we used here is presented in La Mura et al. (2007). Therefore all analysis and
discussions in this paper do not concern with those narrow lines. 21 of them are classified as NLS1s. A two-component model (a
disk line plus a Gaussian component) to fit the profile of the broad
lines has been presented by Popovi\'c et al. (2008) and Bon et al.
(2006). We first tried this model. However, the fitting is not
acceptable statistically for most of the objects in this sample.
Instead, a double Gaussian component model can fit the profile very
well.

First, we fit H$\alpha$ and H$\beta$ lines freely with two Gaussian
components, i.e., a VBGC and an IMGC; a wavelength deviation of 10
$\AA$ from their rest frame wavelength is allowed in the fitting for
the central value of the two components. When the IMGC is
statistically insignificant, we refit the spectrum with just one
Gaussian component. Our results are generally consistent with the
results reported by Mullaney \& Ward (2008); they decomposed nine
H$\alpha$ lines, three of which only need one Gaussian component.
The decomposition of H$\alpha$ is thus quite straight forward.
However, it is more complicated to decompose the H$\beta$ lines,
because of the contamination of a broad line He I at 4922 $\AA$ and
Fe II on the red wing of H$\beta$ (V\'eron et al. 2002; Mullaney \&
Ward 2008). We thus add a Gaussian component with centers ranging
from 4922 $\AA$ to 4940 $\AA$ when necessary. The profile of the
FeII is not important here. Due to the above complication and the
relative weakness of the H$\beta$ lines, sometimes there are
relatively large uncertainties in the H$\beta$ line decomposition.

Second, with the result of the first step, to every H$\alpha$ and H$\beta$ line, we get
the width of their VBGCs, $V_{\rm{H}\alpha}$ and
$V_{\rm{H}\beta}$, respectively. If $V_{\rm{H}\alpha}
>V_{\rm{H}\beta}$, we refit the H$\alpha$ by requiring $|V_{\rm{H}\alpha}-V_{\rm{H}\beta}|\leq0.1 V_{\rm{H}\beta}$
 and fit
the corresponding H$\gamma$ by also requiring
$|V_{\rm{H}\gamma}-V_{\rm{H}\beta}|\leq0.1 V_{\rm{H}\beta}$. Conversely, if
$V_{\rm{H}\alpha} <V_{\rm{H}\beta}$, we refit the H$\beta$ by requiring
$|V_{\rm{H}\alpha}-V_{\rm{H}\beta}|\leq0.1 V_{\rm{H}\alpha}$
 and fit
the corresponding H$\gamma$ by also requiring
$|V_{\rm{H}\gamma}-V_{\rm{H}\alpha}|\leq0.1 V_{\rm{H}\alpha}$. The
motivation of the above requirement is to ensure the fittings are
physical, i.e., the VBGCs of these three lines are produced from
regions with radius not very different, since $dR=-2VdV$, if the
same gravitationally-bound gas dynamics applies to all three lines.
We do not impose the same radius for the emission line regions for
the three VBGCs, because RM measurements have found different delay
times for different emission lines (e.g., Peterson \& Wandel 1999;
Kaspi et al. 2000).
 The difficulty of fitting H$\gamma$ comes from its weak flux, roughly about
two times weaker than H$\beta$. Since H$\alpha$ has a clear red wing, it also helps to
reduce the influence of Fe II contamination in H$\beta$ fitting.

\begin{figure}
\begin{center}
  \includegraphics[angle=0,scale=.6]{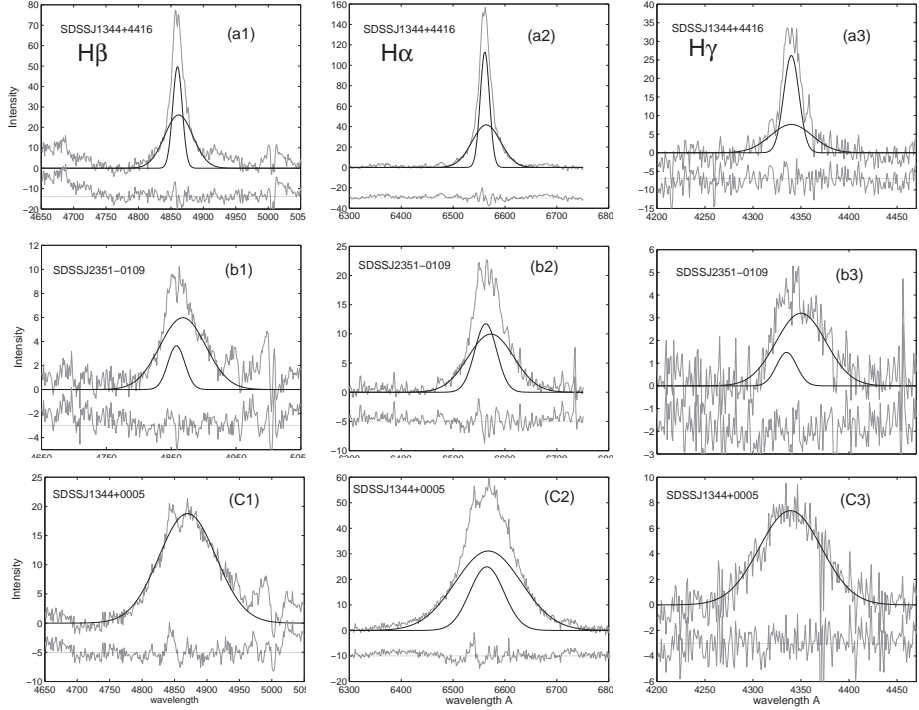}
  \caption{Decomposition examples of broad H$\beta$, H$\alpha$ and H$\gamma$ lines. From (a) to (c), FWHM of the line
  increases, and the IMGC becomes weaker and may disappear sometimes, e.g., in subpanels c1 \& c3.
  Note that for in subpanel c2, the IMGC of H$\alpha$ line is still detectable, in contrast to
  H$\beta$ and H$\gamma$ lines, because the IMGC of H$\alpha$ is normally much stronger than
  H$\beta$ and H$\gamma$ lines (see Fig.\ref{fig:WR} for line intensity ratios between these three lines).)}\label{fig:Hlines}
\end{center}
\end{figure}

A single Gaussian component model is not acceptable for the majority
of the sources in this sample. This is consistent with the results
of Collin et al. (2006), who found that most broad line profiles of
AGNs are non-Gaussian and deviations from a Gaussian profile are
correlated with the properties of AGNs. Almost all of the spectra
(88/90 for H$\alpha$ and H$\beta$, 74/90 for H$\gamma$) can be
fitted very well with this double Gaussian components model. For
this reason, model lines based on two Gaussian components are used
to measure the FWHM of the whole line (e.g., Greene \& Ho 2005).
Since the error estimate is very complex, we just use the typical
error of 10\% of FWHM, following Greene \& Ho (2005) and Vestergaard
\& Peterson (2006). Those that do not fit well have been excluded
from our analysis. Several typical decomposition examples are
presented in Fig.\ref{fig:Hlines} and Fig.\ref{fig:Habk}. The
decomposition parameters of these objects are presented in the
table.

\begin{figure}
\begin{center}
  \includegraphics[angle=0,scale=0.6]{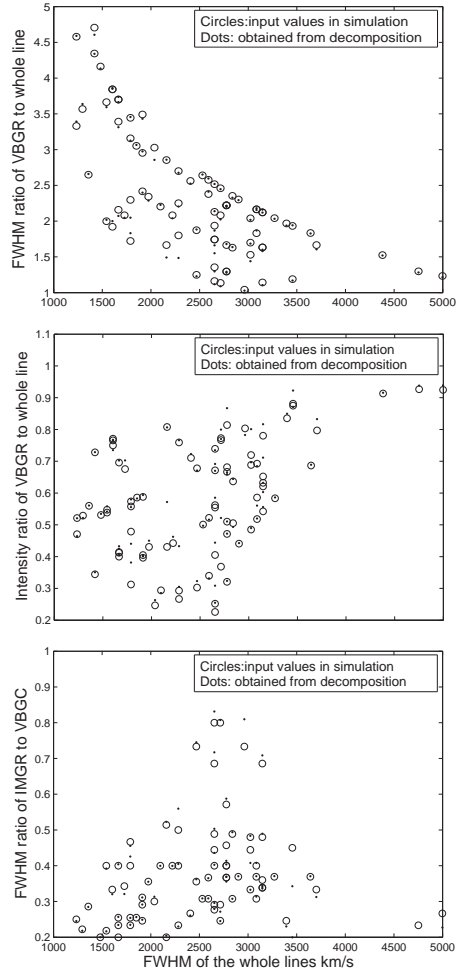}
   \caption{Monte Carlo simulations of spectral decomposition. It can be seen that the decomposed spectral parameters agree with input values quite well, indicating that degenerations in spectral decompositions are not serious.}\label{fig:mont}
  \end{center}
\end{figure}

Since these parameters are somewhat degenerated in the fitting,
Monte Carlo simulations are carried out to examine if this
degeneration affects the reliability of our spectral decompositions.
As shown in Fig.\ref{fig:mont}, 80 simulated spectra are produced,
with each one composed of an IMGC and a VBGC; each parameter of
these Gaussian components is a random choice from a corresponding
parameter group which covers the similar parameter range in our
sample. Proper noise is also added to each simulated spectrum. We
then decompose these simulated spectra and obtain the fitted
parameters. Fig.\ref{fig:mont} shows that the fitted parameters do
not deviate from the initially set parameters significantly.

\section{Analysis of H$\beta$, H$\alpha$ and H$\gamma$ lines}

\subsection{H$\beta$ lines} \label{sec-preamble}

\begin{figure}
\begin{center}
  \includegraphics[angle=0,scale=0.6]{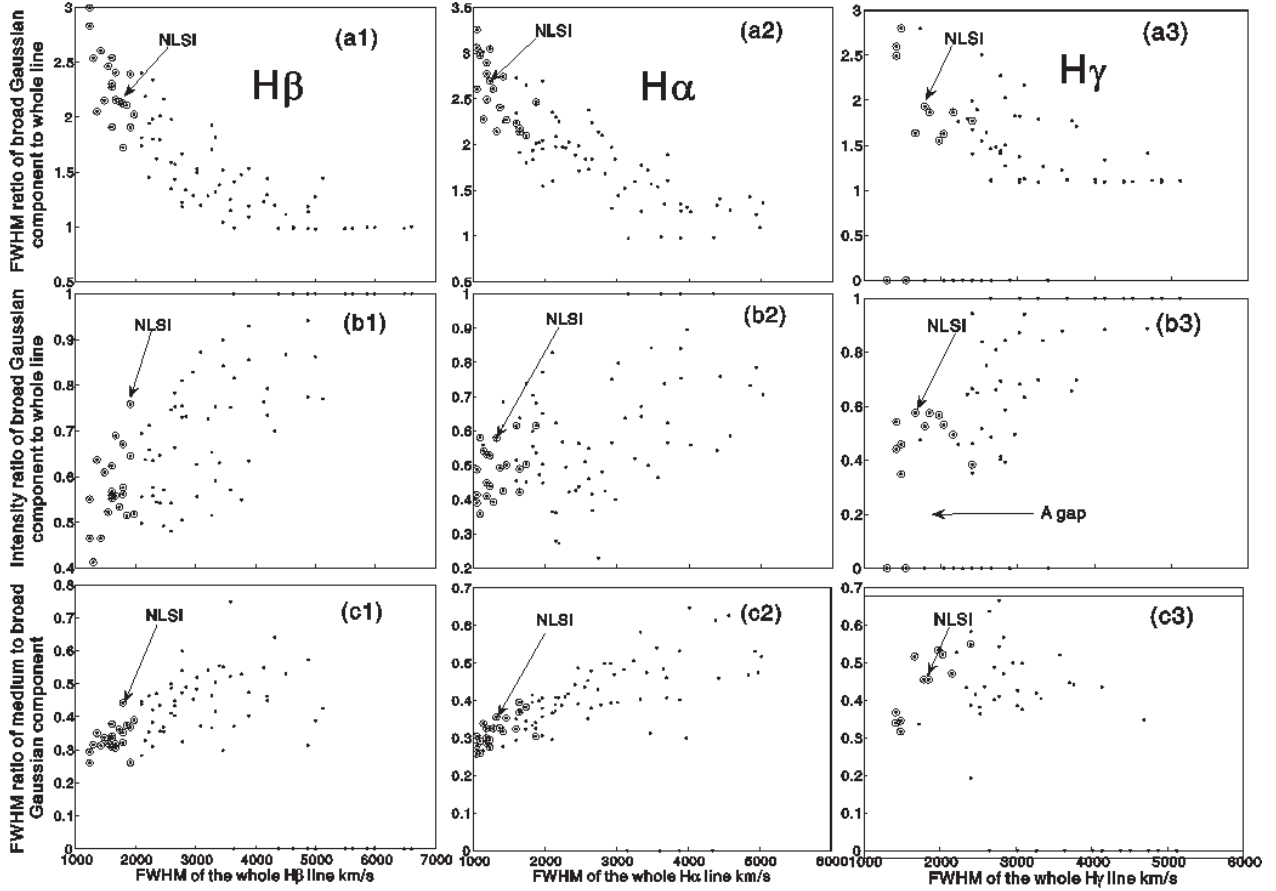}
  \caption{Statistical analysis of broad H$\beta$, H$\alpha$ and H$\gamma$ lines. With the
   FWHM increasing, FWHM ratio of VBGC
    to the whole line becomes smaller and finally reaches
    unity, the intensity ratio of the VBGC
    to the whole line becomes larger and finally also reaches unity, and
  the IMGC becomes broader and weaker. The uncertainty is roughly as large as the dispersion.}\label{fig:hstatistics}
  \end{center}
\end{figure}

First, we focus on H$\beta$ lines; several examples are shown in Fig.\ref{fig:Hlines}
(a1, b1 \& c1) and Fig.\ref{fig:Habk} (b). The statistical analysis is presented in
Fig.\ref{fig:hstatistics} (a1, b1 \& c1). The VBGC has a much
larger FWHM than the whole line when FWHM of the whole line is small. The FWHM ratios
range from 1.5 to 4 in the NLS1 group which lie on the left of
Fig.\ref{fig:hstatistics}(a1). The intensity ratio of the VBGC
to the whole line is about 0.6 in these objects. This means that for NLS1s the
IMGC has an intensity comparable to the VBGC, and
consequently the FWHM of the whole line of NLS1s is dominated by the IMGC. With FWHM increasing, the IMGC becomes
weaker and finally disappeared, which can be seen from Fig.\ref{fig:hstatistics}(b1),
in which the intensity ratio of the VBGC to the whole line
reaches unity when FWHM of the whole line reaches about 5000 km$\cdot$s$^{-1}$.
Naturally, the FWHM ratio of the VBGC to the whole line also
reaches unity. The whole line can thus be simply described by a single VBGC and the VBGC becomes totally dominant. This
trend agrees with that obtained by Hu et al. (2008), who have shown that most of the
AGNs with very large black hole masses (normally with largest FWHM) do not need two
components for their broad H$\beta$ lines. Our result also agrees with Collin et al.
(2006), who found that NLS1s seem to have a prominent narrower component (the
IMGC identified in our work) on top of a broad Gaussian
profile (the VBGC identified in our work).

In Fig.\ref{fig:hstatistics}(c1), FWHM ratio of the IMGC to
the VBGC becomes larger with increasing FWHM. It ranges from $0.2$
to $0.6$, which is another reason for the FWHM of the whole line reaching the FWHM of
the VBGC. It indicates that the IMGC
also becomes broader when it becomes weaker. The points lie on $y=0$ of the bottom
picture represent the objects which have only a single Gaussian component. The three
points with maximum FWHM (but still with two Gaussian components) fall off the trend
slightly, due to perhaps the large uncertainties in fitting when the intensity of the
IMGC becomes very low.

Compared to Fig.2, the correlations in Fig.3 are much tighter,
suggesting that the two-components decomposition is physically
meaningful, not just a mathematical convenience. Moreover, the
systematical evolution show in Fig.3 suggests that the two
components are physically distinct, and both components responses to
a common source.
\subsection{H$\alpha$ lines} \label{sec-preamble}

Fig.\ref{fig:Hlines} (a2, b2 \& c2) and Fig.\ref{fig:Habk} (a) are several
decomposition examples for H$\alpha$ lines. They behave similarly when FWHM is small.
However when FWHM is large, the IMGC is very broad and
strong. Statistical analysis is shown in Fig.\ref{fig:hstatistics} (a2, b2 \& c2) in a
similar manner to H$\beta$ lines. The obvious difference is that the number of points
near unity in panel (b2) is less than that in panel (b1) for H$\beta$ lines; only a
weakly increasing trend can be seen in panel (b2). This means that the IMGC is still strong even when FWHM reaches about 5000 km$\cdot$s$^{-1}$.

\subsection{H$\gamma$ lines} \label{sec-preamble}

\begin{figure}
\begin{center}
  \includegraphics[angle=0,scale=.5]{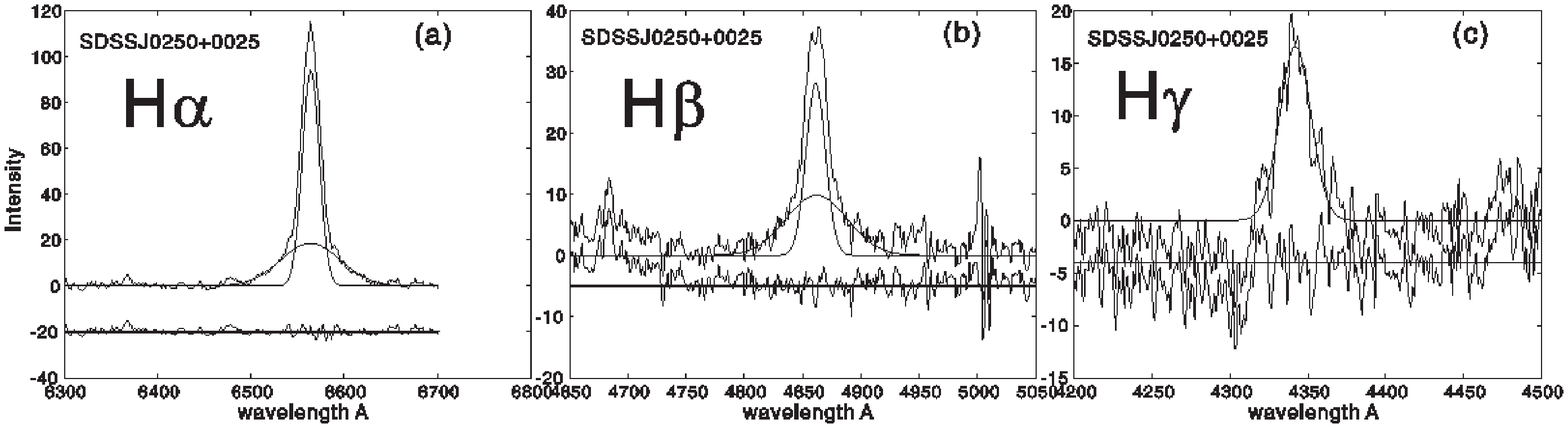}
  \caption{Decompositions of H$\alpha$, H$\beta$, H$\gamma$ of SDSSJ0250+0025.
   Its H$\beta$ and H$\alpha$ lines behave similarly as SDSSJ1344+4416 shown in Fig.1(a1) and Fig.1(a2), respectively.
   The H$\gamma$ line lost its VBGC, which can be caused by confusion
   in the continuum subtraction, when it is very weak and broad.}\label{fig:Habk}
\end{center}
\end{figure}

H$\gamma$ lines behave similarly to H$\beta$ lines as shown Fig.\ref{fig:Hlines} (a3,
b3 \& c3) and Fig.\ref{fig:Habk} (c); when FWHM is very large, only a single Gaussian
component (the VBGC) is required to fit the profile. However,
when FWHM is very small, a large number of H$\gamma$ lines loose the VBGC, i.e, only the IMGC is required to fit the line,
because most of their FWHM are very close to the IMGCs of
their H$\beta$ lines. The points lie on the x-axis in Fig.\ref{fig:hstatistics} (b3)
are the cases when the H$\gamma$ lines are described by a single IMGC; an example is presented in Fig.\ref{fig:Habk} (c). However some of the
single Gaussian component H$\gamma$ lines have FWHM neither close to the VBGC nor close to the IMGC of the
corresponding H$\beta$ lines, therefore we exclude these points in our analysis. The
loss of the VBGCs may be caused by the weak intensity of
H$\gamma$, since its VBGC is particularly weak when FWHM is
small, which can cause confusion in the continuum subtraction. The gap pointed by an
arrow in Fig.\ref{fig:hstatistics} (b3) also reveals a sudden decrease of the intensity
of the broad Gaussian component when it is very weak.

\subsection{FWHM evolution with luminosity}
\begin{figure}
\begin{center}
  \includegraphics[angle=0,scale=0.6]{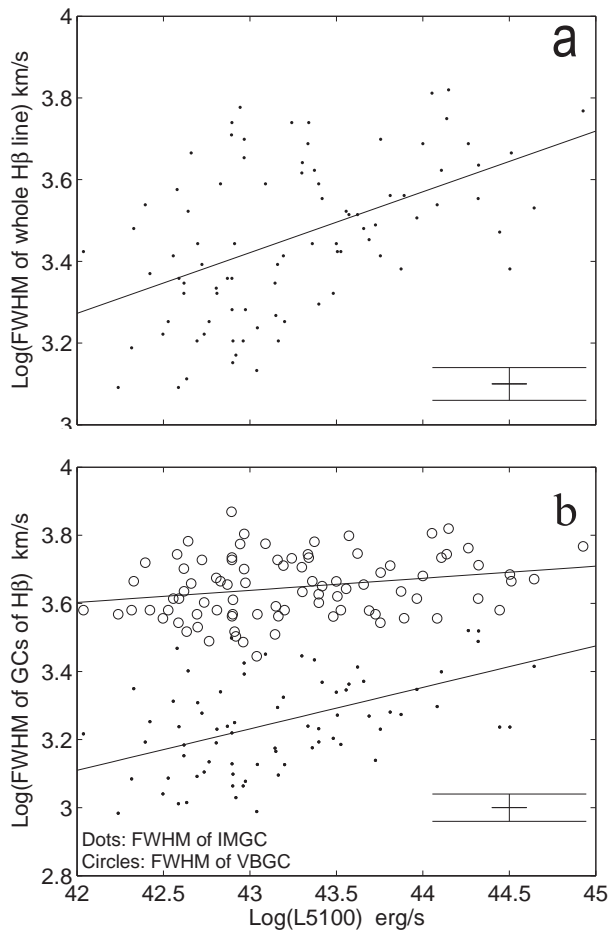}
   \caption{Upper panel shows that FWHM of the whole H$\beta$ line increase with luminosity of the source. Lower panel is the correlation between FWHM
   of a single Gaussian component (VBGC/IMGC) and luminosity. The sum squared error (SSE) of the L5100-FWHM fitting in (a) is 2.34 (88 points),
    SSE of the L5100-FWHMb fitting in (b) is 0.66 (88 points) and SSE of the L5100-FWHMi fitting is 1.07 (74 points).
    The correlation of a single Gaussian component' FWHM and luminosity is tighter than that of the whole line. The typical uncertainty is plotted in the corner}\label{fig:L-FWHM}
  \end{center}
\end{figure}

Previous works suggested that a highly significant correlation
exists between FWHM(H$\beta$) and source luminosity (e.g., Joly et
al. 1985; Corbett et al. 2003). More recent work with a larger
sample which spans a FWHM range from 1000 to 16000 km s$^{-1}$
showed that the correlation is not so strong but still statistically
significant; the correlation in this large sample is likely driven
by the minimum FWHM trend (Marziani et al 2009).
Fig.\ref{fig:L-FWHM}a presents this correlation in our sample; the
correlation is more obvious in this smaller sample than in Marziani
et al (2009), and it is very similar to the correlation in Joly et
al (1985). The extremely broad H$\beta$ (FWHM $>$ about 7000 km
s$^{-1}$) may have a rather different line profile and may originate
from a different physical region from the less broad H$\beta$ (e.g.,
Strateva et al. 2003; Eracleous et al. 2003), therefore the analysis
with only mediate broad (FWHM $<$ about 7000 km s$^{-1}$) H$\beta$
with the same line profile seems to be more meaningful.
Fig.\ref{fig:L-FWHM}b shows that the logarithm of FWHM of the VBGC
or IMGC increases with the logarithm of luminosity in a more linear
way than FWHM of the whole line. This may indicate that FWHM of VBGC
or IMGC is more physical than FWHM of the whole line. On the other
hand, Fig.\ref{fig:L-FWHM}b also shows that FWHM of IMGC increases
with luminosity much faster than VBGC; the two components has a
trend to merge into one. This trend will be confirmed in another way
in section 4.1. Therefore FWHM evolution also supports the picture
that the broad line region is composed of two physically distinct
regions.

%% The equation environment wil produce a numbered display equation.

\section{Dynamics and physical properties of the two emission regions}

\subsection{Evolution of the two emission regions} \label{sec-preamble}
We assume that the two Gaussian components come from two distinct
regions, and both of the these two regions (VBLR and IMLR) are
bounded
by the central black hole's gravity. We therefore have\\
\begin{equation}
f_1^2\frac{V_{\rm{ VBLR}}^{2}}{R_{\rm{ VBLR}}}=\frac{GM_{\rm{
BH}}}{R_{\rm{ VBLR}}^{2}},
\end{equation}
and\\
\begin{equation}
f_2^2\frac{V_{\rm{ IMLR}}^{2}}{R_{\rm{ IMLR}}}=\frac{GM_{\rm{
BH}}}{R_{\rm{ IMLR}}^{2}}.
\end{equation}
Equation (1) and (2) lead to\\
\begin{equation}
R_{\rm{ IMLR}}=(\frac{f_1V_{\rm{ VBLR}}}{f_2V_{\rm{
IMLR}}})^{2}R_{\rm{ VBLR}}=C_0(\frac{V_{\rm{ VBLR}}}{V_{\rm{
IMLR}}})^{2}R_{\rm{ VBLR}},\\
\end{equation}
where
\begin{equation}
C_0=(\frac{f_1}{f_2})^2.\\
\end{equation}
We assume that $C_0$ is a constant for all sources, because currently we cannot determine the exact geometry and velocity
 distribution of gases in the IMLR. The validity of this assumption and the exact value of $C_0$ needs to be determined with
 further studies of more observation data. For convenience we simply redefine
$R_{\rm{ IMLR}}=R_{\rm{ IMLR}}/C_0$ for the rest of this paper,
i.e., we only consider the special case when $C_0=1$ (see section
5.2 for a tentative support to this rather arbitrary choice).

RM is used to measure radius of the BLR, which is related to an
AGN's continuum luminosity by the
following empirical Radius-Luminosity relation (Kaspi et al. 2005):\\
\begin{equation}
\log{\frac{R_{\rm{ BLR}}}{\rm{lt-days}}}=(0.69 \pm 0.05)
\log{\frac{\lambda L_{\lambda}(5100 \AA)}{\rm{erg}\cdot
\rm{s}^{-1}}}-29.0\pm 2.2,
\end{equation}
or the starlight-corrected R-L relation (Bentz et al. 2006): \\
\begin{equation}
\log{\frac{R_{\rm{ BLR}}}{\rm{lt-days}}}=(0.518 \pm 0.039)
\log{\frac{\lambda L_{\lambda}(5100 \AA)}{\rm{erg}\cdot
\rm{s}^{-1}}}-21.2\pm 1.7.
\end{equation}
We use equation (6) to calculate $R_{\rm{ BLR}}$ throughout this
work, unless indicated otherwise. Equation (5) is only used in
Section 4.4 where a comparison is made.

\begin{figure}
\begin{center}
  \includegraphics[angle=0,scale=.7]{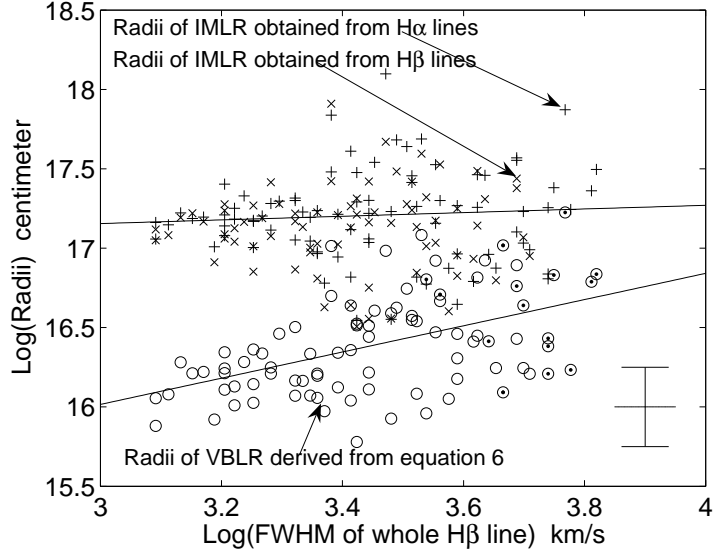}\\
  \caption{Correlation of the radii of IMLR and VBLR with the FWHM of the whole H$\beta$ line. Radii are plotted in logarithm scale. Circles are the VBLR radius,
  ``$\times$'' represent IMLR radii derived from H$\beta$ lines and  ``$+$'' are obtained from
  H$\alpha$ lines. Circles filled with dots represent the objects missing the IMGC. Least-absolute-residuals fitting results are also plotted:
   for IMLR $\log{R}= 0.115(\pm0.3)\log{\rm{FWHM}}+16.81(\pm1.1)$, and for VBLR
        $\log{R}= 0.8253(\pm0.27)\log{\rm{FWHM}}+13.54(\pm1)$. It is interesting to note that the radius of IMLR is not correlated
        with the FWHM of the whole H$\beta$ line. The typical uncertainty is plotted in the corner.}
   \label{fig:RRF}
  \end{center}
\end{figure}

Because RM is used to measure radius of BLR, one might expect that
it would normally measure the radius of the innermost emission-line
region, i.e., the VBLR in our model (see discussion on this issue in
section 5.2). Therefore, the radius calculated from $R_{\rm{ BLR}}
\sim$L5100 relation is taken as $R_{\rm{ VBLR}}$. Consequently, we
can calculate the radius of IMLR from equation (3). Both H$\alpha$
and H$\beta$ lines can be used to do this calculation. Because they
seem to behave slightly differently in the line evolution, $R_{\rm{
IMLR}}$ is obtained from H$\alpha$ and H$\beta$ lines separately.
Fig.\ref{fig:RRF} is the evolution of $R_{\rm{ VBLR}}$ and $R_{\rm{
IMLR}}$ with FWHM. IMLR radii derived from H$\alpha$ and H$\beta$
lines have only subtle differences. The IMLR radius just varies
around a constant with increasing FWHM, i.e. does not have a
systematic increase or decrease. On the other hand, VBLR radius
becomes larger with increasing FWHM. VBLR and IMLR are clearly
separated with FWHM$<2500$ km$\cdot$s$^{-1}$ as we can see in
Fig.\ref{fig:RRF}. The circles filled with dots represent the
objects whose lines are fitted well with a single (very) broad
Gaussian component; they appear in the Fig.\ref{fig:RRF} with
largest FWHM.

The evolution of $R_{\rm{ VBLR}}$ and $R_{\rm{ IMLR}}$ with
luminosity and black hole mass are shown in Fig.\ref{fig:RR}. The
radius of the IMLR increases more slowly than VBLR with black hole
mass and luminosity. The two emission regions have a trend to merge
into one with higher luminosity or larger black hole mass; it is
consistent with the emission lines' behavior as shown in section
3.4. The points which represent objects with zero IMGCs also mainly
lie on the large mass and large luminosity side. The slopes of these
relations will be discussed in Section 4.4. Although most of the
H$\alpha$ lines still need two Gaussian components to fit, there are
three of them which only need one Gaussian component in our sample.
The two Gaussian components in H$\alpha$ lines also show a trend to
merge into one single Gaussian component with FWHM becoming larger,
though the trend is not as obvious as for H$\beta$ lines. Once again
the systematically different trends of evolution of $R_{\rm{ VBLR}}$
and $R_{\rm{ IMLR}}$ are consistent with our assumption that they
are physically two distinct emission regions.

\subsection{Balmer decrement in different Gaussian components and stratified geometry of the IMLR} \label{sec-preamble}

It is possible that density, temperature and geometric shape of the
two regions (VBLR and IMLR) are different, then the collisional
effect and different ionization energies of the three Balmer lines
may cause different Balmer decrement in the two regions. Here we
show the Balmer decrement of the whole line, VBGC and IMGC in
Fig.\ref{fig:WR}. There are two sources whose $\rm{H}\alpha$ or
$\rm{H}\beta$ or both do not fit well with the double Gaussian
model; these two points are excluded in Fig.\ref{fig:WR}a when
calculating the intensity ratio of $\rm{H}\alpha$ to $\rm{H}\beta$.
Similarly, there are sixteen sources whose $\rm{H}\beta$ or
$\rm{H}\gamma$ or both do not fit well; these points are excluded
when calculating the intensity ratio of $\rm{H}\beta$ to
$\rm{H}\gamma$. The points excluded in Fig.\ref{fig:WR}a are also
excluded in Fig.\ref{fig:WR}b and Fig.\ref{fig:WR}c. In addition,
fifteen sources' $\rm{H}\gamma$ do not have broad Gaussian; these
points do not appear in intensity ratio of H$\beta$ to
$\rm{H}\gamma$ in Fig.\ref{fig:WR}b. Fourteen sources'
$\rm{H}\alpha$ or $\rm{H}\beta$ or both only need a VBGC to fit
(i.e., they do not have IMGC), so these sources are not included in
Fig.\ref{fig:WR}c when calculating the $\rm{H}\alpha$ to
$\rm{H}\beta$ ratio. Similarly 22 sources' $\rm{H}\beta$ or
$\rm{H}\gamma$ or both only have VBGC, and are thus excluded in
Fig.\ref{fig:WR}c when calculating the $\rm{H}\beta$ to
$\rm{H}\gamma$ ratio. In summary, there are 88, 88 and 76 points
used to calculating H$\alpha$ to H$\beta$ ratio in
Fig.\ref{fig:WR}a, b and c, respectively; correspondingly, there are
74, 59 and 52 points in H$\gamma$ to H$\beta$ ratio.

In VBLR, H$\alpha$/H$\beta$ is 2.54 and H$\gamma$/H$\beta$ is 0.37,
which is consistent with a pure H model with \mbox{T=15000 K},
Ne=$10^{10}$ cm$^{-3}$ (Krolik et al. 1978). In IMLR,
H$\alpha$/H$\beta$ is 4.78, which is much higher than that in VBLR,
consistent with previous results (Netzer \& Laor 1993; Mullaney \&
Ward 2008). These can be caused by collisional effect which can lead
to an effective emission increase for H$\alpha$ but not for H$\beta$
or H$\gamma$ (e.g. Osterbrock 1989). This result suggests a higher
gas density in IMLR. It has also been argued that dust in the
emission region can cause higher Balmer decrement (Binette et al.
1993). Therefore, it also suggests that IMLR may be contaminated by
dust. On the other hand, the H$\gamma$ to H$\beta$ ratio is also
slightly higher in IMLR (0.41) than that in VBLR (0.37). As H$\beta$
and H$\gamma$ are produced from similar process, the slightly higher
H$\gamma$ to H$\beta$ ratio is not well understood.

% stratified geometry of IMLR.
In Fig.\ref{Cor_lines1}, we show the correlations between the first
three Balmer lines for the IMGC, i.e., between the H$\alpha$
   and H$\beta$ lines, and between the H$\gamma$
   and H$\beta$ lines. The linear correlations between them indicates that the IMGCs for all these three lines
originate from physically connected regions, even if not exactly
from the same region. In Fig.\ref{Cor_lines2}, we show the
distributions of FWHM differences between the three lines for the
IMGC. The FWHM of H$\alpha$ and H$\gamma$ is offset systematically
by around -200 km s$^{-1}$ and +200 km s$^{-1}$ around that of
H$\beta$, respectively. This suggests a stratified geometry for the
IMLR, where H$\gamma$, H$\beta$ and H$\alpha$ lines are produced at
increasing radii respectively. FWHM of the whole line, H$\alpha$ is
also systematically larger than H$\beta$ (Greene \& Ho 2005, Shen et
al 2008). It is consistent with the RM result (Kaspi et al. 2000)
which showed that radius of BLR obtained from H$\alpha$ is larger
than that of H$\beta$ in most of the sources. Although there is no
difference in the ionization degree of H$\alpha$, H$\beta$ and
H$\gamma$, the above result can be explained that the inner skin of
torus is the IMLR, therefore, it has more dust in the region with
larger radius. There are mainly two processes. First, dust reddening
will make the H$\gamma$ produced in the region much closer to the
torus be extincted most seriously, the H$\beta$ will suffer less
extinction and H$\alpha$ will be least extincted. Therefore the
average radius of the H$\alpha$ region will be larger than H$\beta$
and H$\gamma$. Second, collisional excitation contributes a lot to
H$\alpha$ line but not significantly to H$\beta$ and H$\gamma$ line,
so the region much closer to the torus with a higher density will
produce relatively more H$\alpha$. This also causes the FWHM of
H$\alpha$ to be smaller than H$\beta$ and H$\gamma$.

\subsection{Baldwin effect in Intermediate Gaussian component} \label{sec-preamble}

The relation between the irradiation region and the central
continuum is important, since the emission lines are thought to be
mostly caused by photoionization. Fig.\ref{fig:cw} is plotted based
on the analysis of H$\beta$ lines. Baldwin effect can be seen in the
bottom plot. The Pearson Correlation Coefficient (PCC) is -0.39. We
bootstrap it 100000 times and obtain 10 PCC larger than 0, so PCC is
smaller than 0 at 99.99\% confidence level. Pearson Rank Correlation
Coefficient (PRCC) is -0.48, at 99.9\% confidence level smaller than
0 using the same method as the confidence level calculation of PCC.
However no Baldwin effect is seen in the top (Pearson correlation
coefficient is -0.1) or in the middle plot (Pearson correlation
coefficient is 0.1). It means that the Equivalent Width (EW) of the
IMGC becomes smaller when the AGN becomes brighter, whereas the EW
of the VBGC does not change with the continuum luminosity. This
phenomenon has also been seen by Marziani et al. (2009). Since EW
reflects the covering factor of the emission region, the above
result supports the scenario that the VBLR is nearly spherical, but
the IMLR has a flattened geometry. Therefore in the absence of other
knowledge of other parameters of the gas, we can use the higher EW
as a supportive (albeit not conclusive) evidence for a higher
covering factor. There is also evidence that the dust torus has a
smaller covering factor with higher luminosity (Wang et al. 2005).
We therefore suggest that the cause of the slight Baldwin effect of
IMLR can be the same, i.e., the innermost region of the dusty torus
can be sublimated away by the strong irradiation of the AGN.
Therefore we can infer that IMLR and VBLR have different geometries.

\subsection{Location and Geometry of IMLR} \label{sec-preamble}

\begin{figure}
\begin{center}
  \includegraphics[angle=0,scale=.7]{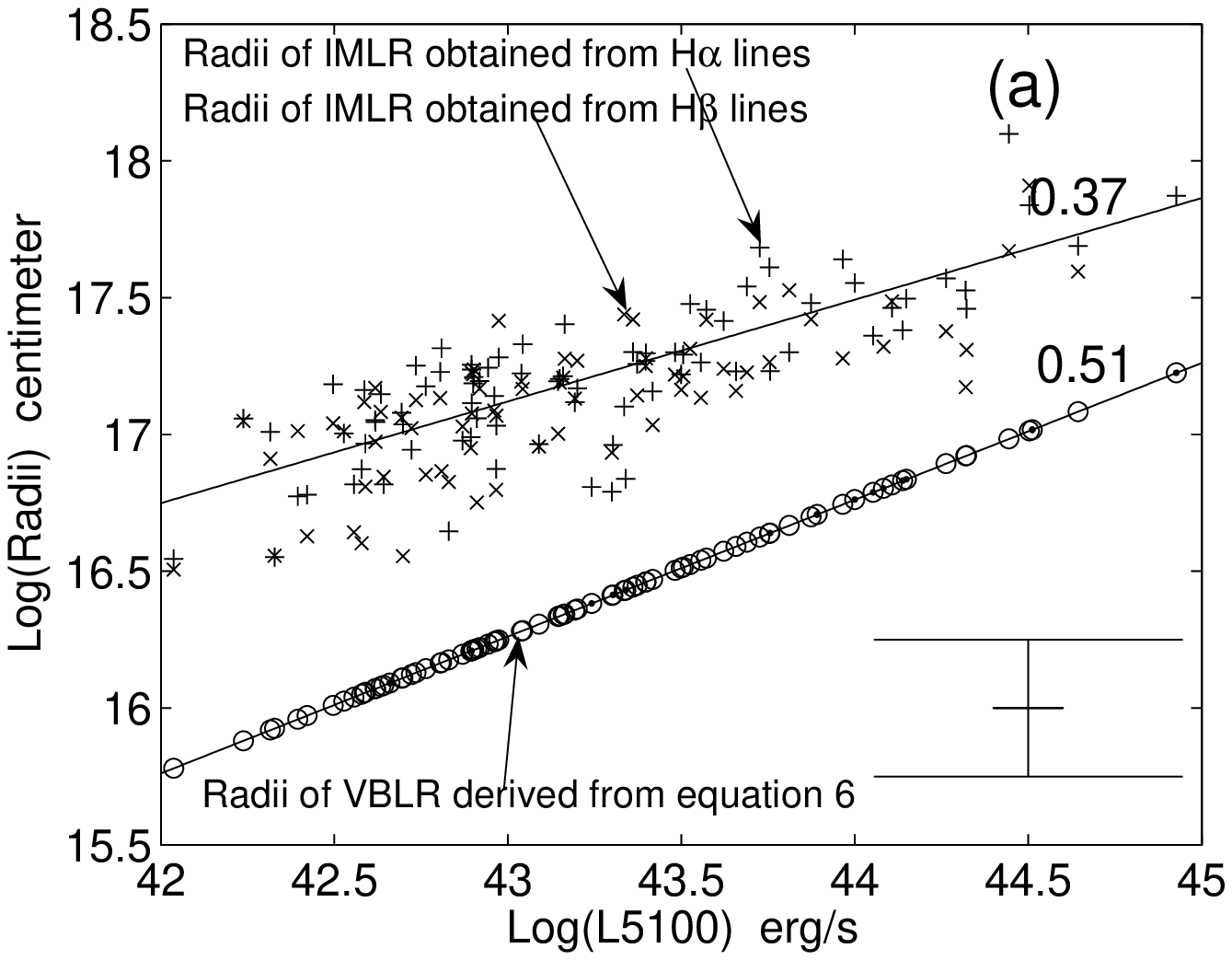}\\
  \includegraphics[angle=0,scale=.7]{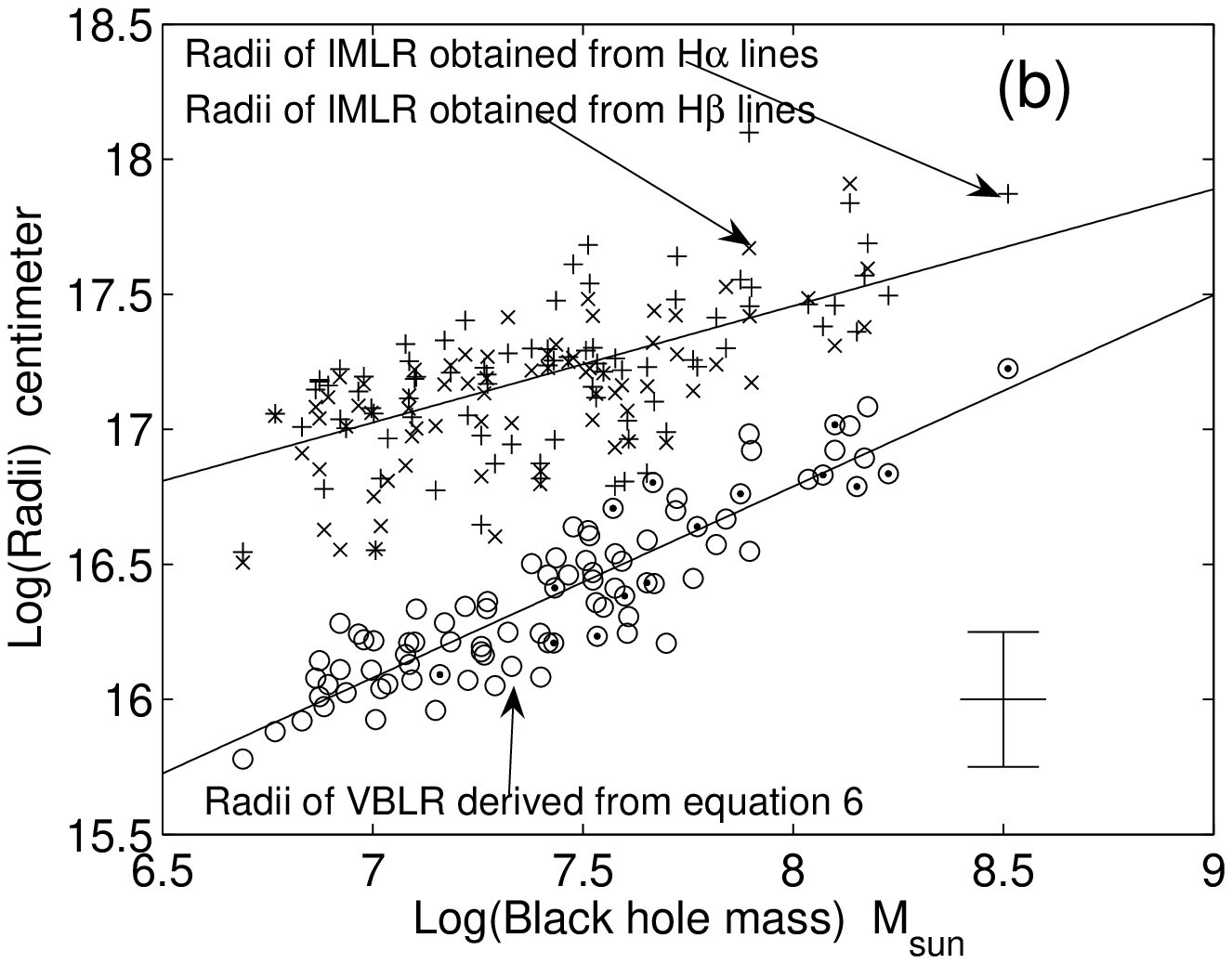}\\
  \caption{Correlation of the radii of IMLR and VBLR with black hole mass. Circles are the VBLR radius, $\times$ represent IMLR radius derived
  from H$\beta$ lines and $+$ from H$\alpha$ lines. Circles filled with dots
  represent the objects missing the IMGC. The radius of the VBLR is obtained from the $R_{\rm{ BLR}} \sim$L5100 (equation 3) relationship on the
  top figure. These correlations supports a scenario of hierarchical evolution of AGNs. The typical uncertainty is plotted in the corner.}\label{fig:RR}
  \end{center}
\end{figure}

\begin{figure}
\begin{center}
  \includegraphics[angle=0,scale=.7]{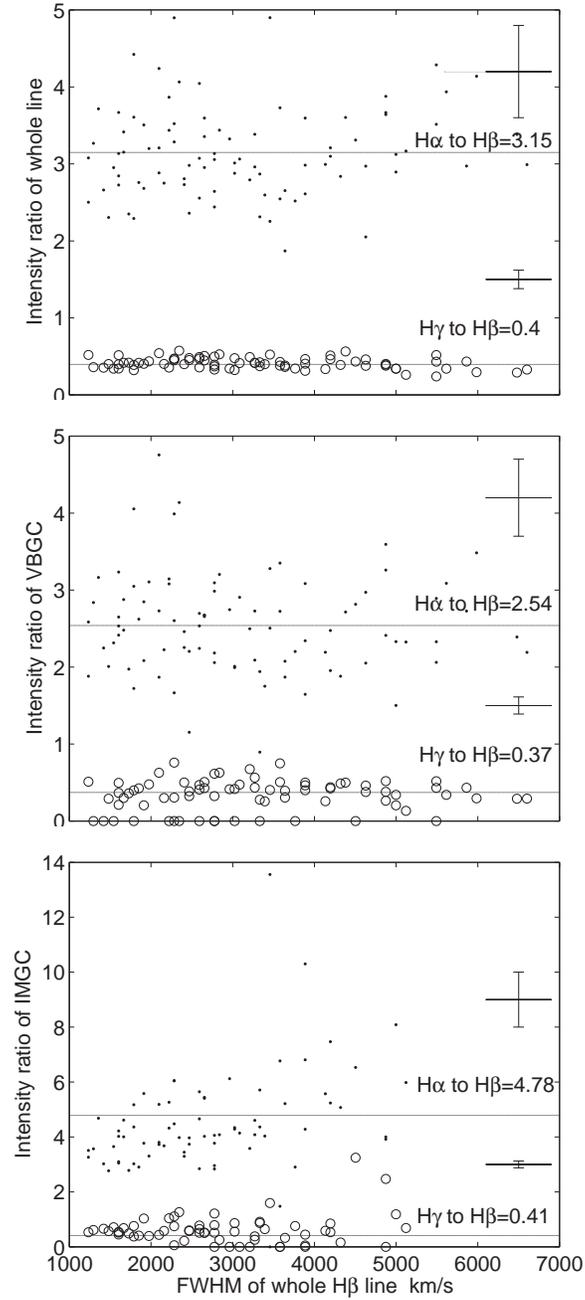}\\
   \caption{ Dots represent intensity ratio of H$\alpha$ to H$\beta$ lines. Circles represent
  H$\gamma$ to H$\beta$ lines. The numbers marked in the figure are the averaged
  values. Typical uncertainty is plotted as a cross.}\label{fig:WR}
  \end{center}
\end{figure}

\begin{figure}
 \begin{center}
   \includegraphics[angle=0,scale=0.7]{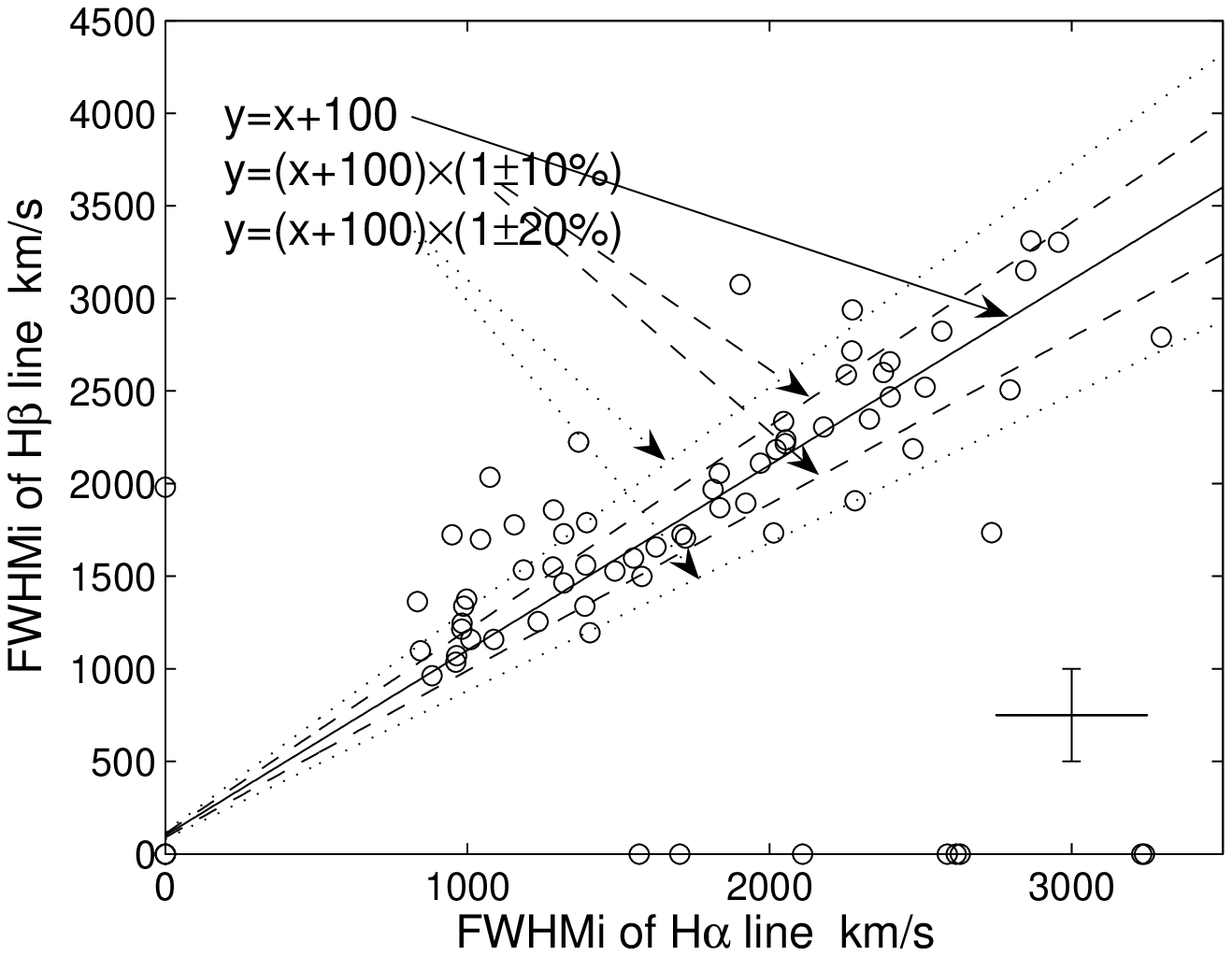}\\
   \includegraphics[angle=0,scale=0.7]{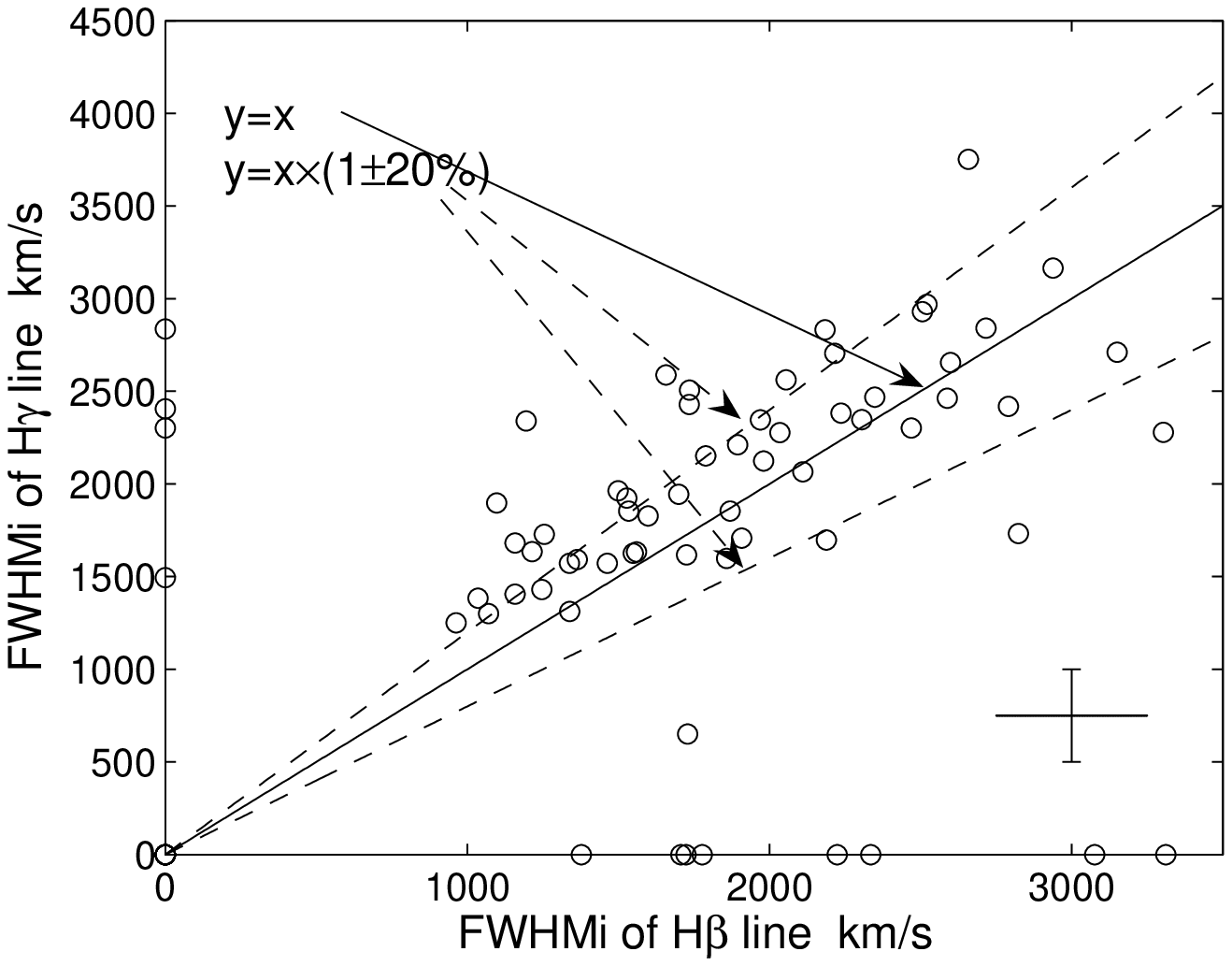}
   \caption{Correlations of between the three lines for the IMGC. {\it Upper panel}: between the H$\alpha$
   and H$\beta$; {\it Lower panel}:  between the H$\gamma$
   and H$\beta$. The linear correlations between them indicates that the IMGCs for all these three lines
originate from physically connected regions, even if not exactly
from the same region. Typical uncertainty is plotted as a
cross.}\label{Cor_lines1}
   \end{center}
 \end{figure}

\begin{figure}
 \begin{center}
   \includegraphics[angle=0,scale=0.7]{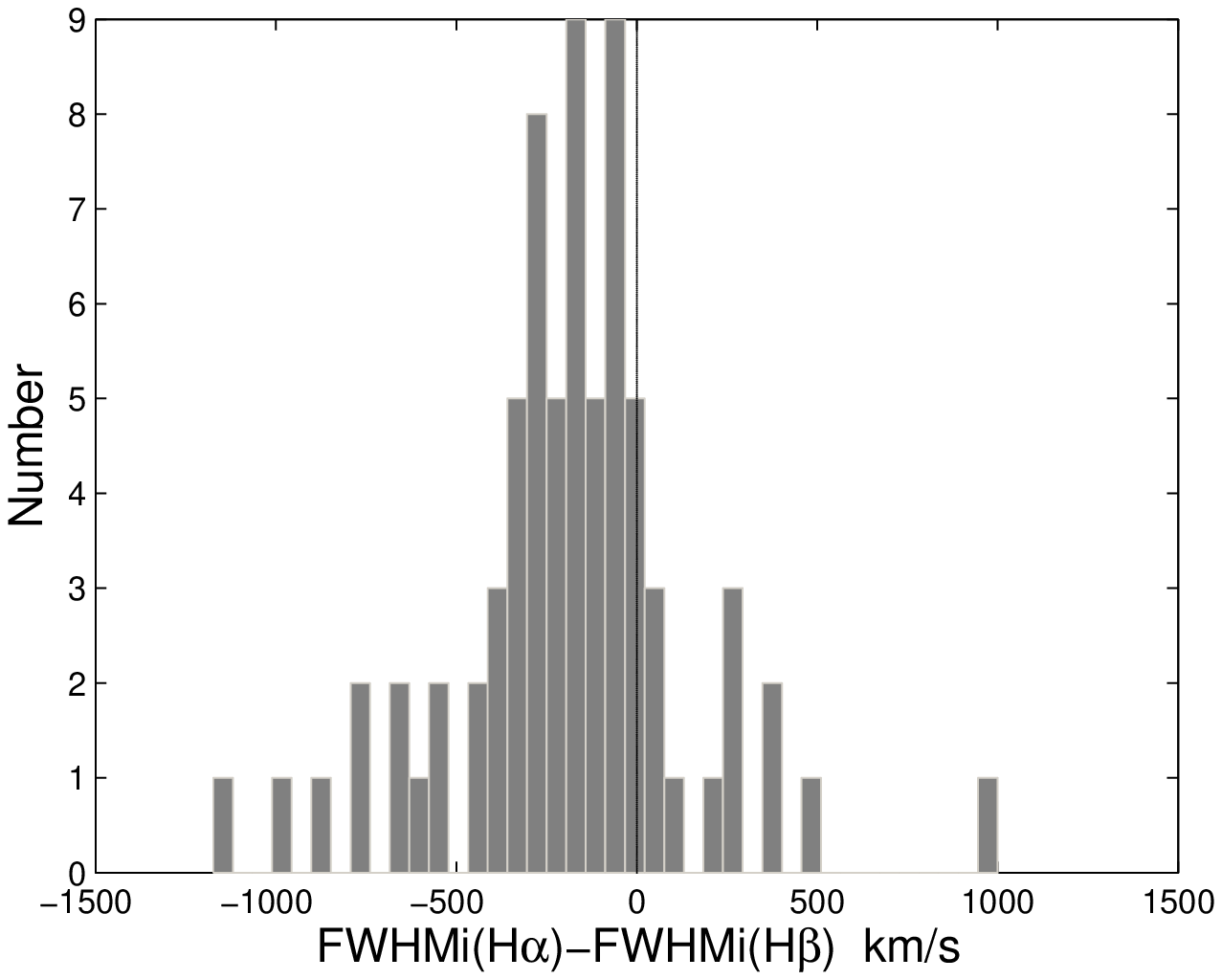}\\
   \includegraphics[angle=0,scale=0.7]{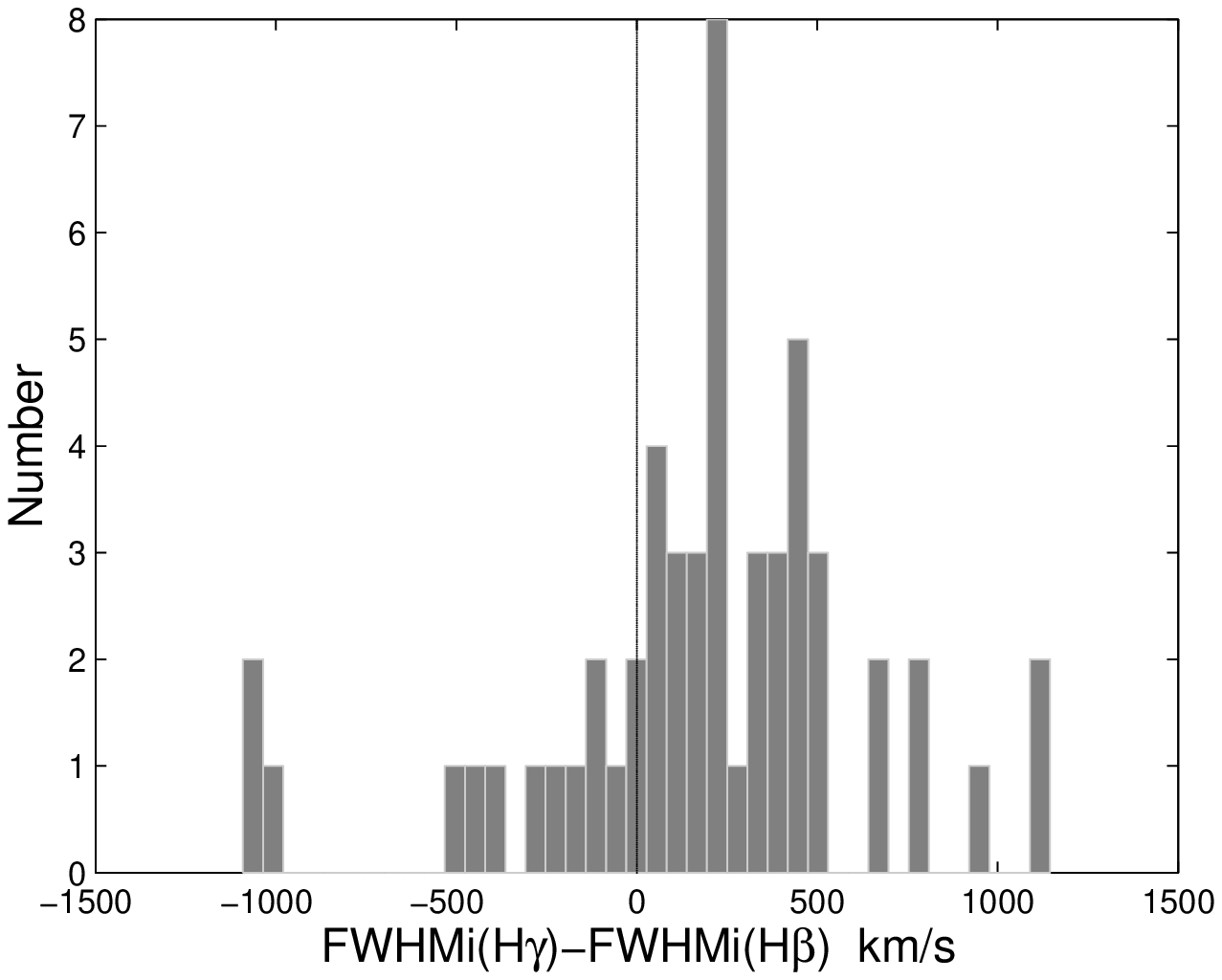}
   \caption{Distribution of FWHM differences between the three lines for the IMGC. {\it Upper panel}: between the H$\alpha$
   and H$\beta$; {\it Lower panel}:  between the H$\gamma$
   and H$\beta$. The FWHM of H$\alpha$ and H$\gamma$ is offset systematically by around -200 km/s and +200 km/s around
   that of H$\beta$, respectively. This suggests a stratified geometry for the IMGC,
   where H$\alpha$, H$\beta$ and H$\gamma$ lines are produced at increasing radii
   respectively.}\label{Cor_lines2}
   \end{center}
 \end{figure}

\begin{figure}
\begin{center}
  \includegraphics[angle=0,scale=.6]{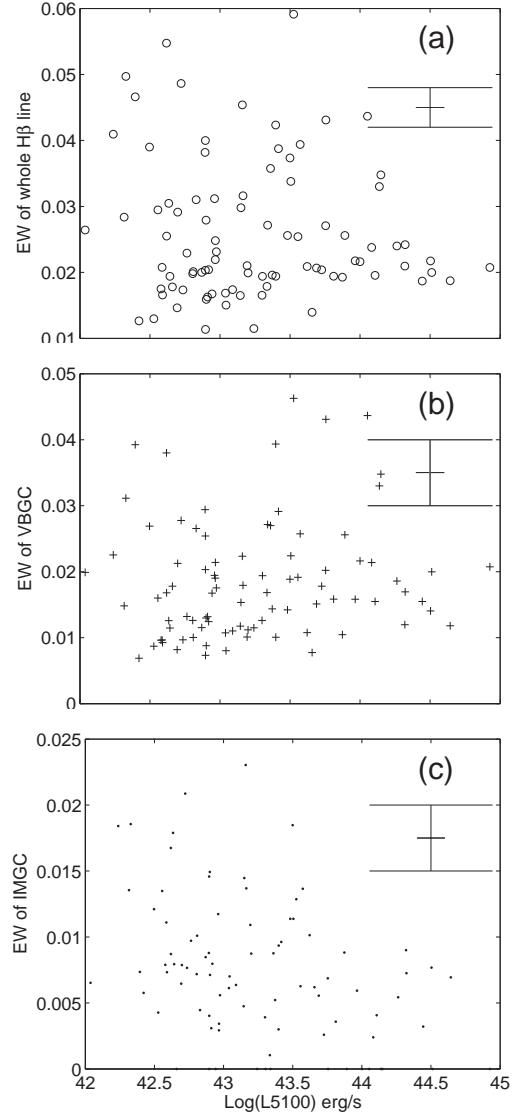}\\
   \caption{Correlation between equivalent width and continuum luminosity, based on the analysis of H$\beta$ lines. Slight Baldwin effect is shown in the IMGC
  (bottom plot) but not on the VBGC. Typical uncertainty is plotted as a cross.}\label{fig:cw}
  \end{center}
\end{figure}

Based on the above analysis, a picture of IMLR has emerged. In
comparison with VBLR (the traditionally called ``Broad Line Region"
with a near spherical structure in virialization equilibrium), it
has a larger radius and higher density, contains more dust, is more
flattened, and is thus consistent as being the inner boundary region
of the dust torus of AGN. As shown in Fig.\ref{fig:RR}, the radii of
both VBLR and IMLR are primarily determined by the continuum
luminosity of the AGN. $R_{\rm{ BLR}}$ obtained with equation (6) is
used as the radius of VBLR there. Then $R_{\rm{ IMLR}}\propto
L^{0.37\pm0.06}_{\rm{ opt}}$ is derived from equation (3) as we have
shown in Fig.\ref{fig:RR}(a). The stratified geometry of IMLR, as
inferred from Fig.\ref{Cor_lines2} in section 4.2, is also
consistent with the above picture.

\begin{figure}
\begin{center}
  \includegraphics[angle=0,scale=.7]{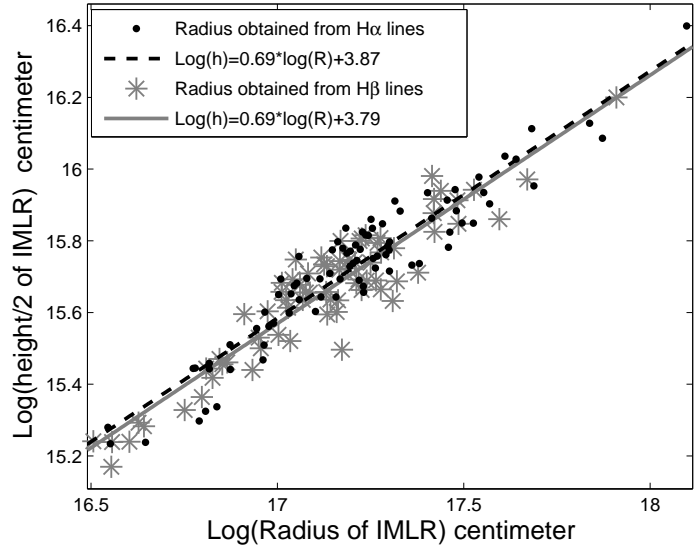}\\
  \caption{Dots and circles represent $h$ obtained from $R_{\rm{ IMLR}}$ based on H$_{\alpha}$ and
  H$_{\beta}$ separately; they have negligible differences. The height of the inner
  torus (IMLR) increases when its radius increases with an index smaller than
  unity.}\label{fig:RHH}
  \end{center}
\end{figure}

The receding velocity of torus with luminosity can be also obtained
from the analysis of type I AGN fraction. A relation between the
covering factor $C$ and $L5100$ can be obtained from Maiolino et al.
(2007): $\log{C}=-0.18\log{(\frac{L5100}{erg/s})}+7.4$, consistent
with Wang et al (2005). If $R_{\rm{ IMLR}}$ is used as the radius of
the inner torus, then $C=4\pi\sin{\theta}=4\pi\frac{h}{h^2+R^2_{\rm{
IMLR}}}$, and the height of the inner torus $h$ can be calculated
from this relation, as shown in Fig.\ref{fig:RHH}. Clearly the
relationship between $h$ and $R_{\rm{ IMLR}}$ can be fitted with a
power-law form. Since $C_{1}$ affects the index very weakly in this
fitting, we just let $C_{1}=0$ in Fig.\ref{fig:RHH}. The height of
the inner torus increases when its radius increases with the
relation $h\propto R^{0.69}\propto L^{0.25}$, which controls the
geometry of torus.

 RM based on infrared emission is
consistent with $R_{\rm{torus}} \propto L^{0.5}_{\rm{ opt}}$
(Suganuma et al. 2006), which agrees with the prediction of a toy
model for the sublimation process (Krolik \& Begelman 1988).
However, Suganuma et al. (2006) did not provide the index' error of
$R_{\rm{torus}}-L_{\rm{ opt}}$ relation. We refit the data with
error and the best fitting is $R_{\rm{torus}}\propto
L^{0.46\pm0.09}_{\rm{ opt}}$, statistically consistent with our
relation between IMLR radius and luminosity ($R_{\rm{ IMLR}}\propto
L^{0.37\pm0.06}_{\rm{ opt}}$).

 However, the radius of IMLR that we calculated here is
strongly dependent on the radius of VBLR, i.e., dependent on the
$R_{\rm{ BLR}}\sim L5100$ relation. If equation (5) is used to
calculate the radius of BLR which is used as the radius of VBLR
here, we would obtain $R_{\rm{ IMLR}}\propto L^{0.53\pm 0.07}$,
which is also consistent with the infrared RM result of
$R_{\rm{torus}}\propto L^{0.46\pm0.09}_{\rm{ opt}}$. However, we
prefer equation (6) because it has taken into account the correction
of starlight (Bentz et al. 2006). Equation (6) is also consistent
with a simple radiation pressured dominated VBLR where
$L/R^2=$constant.

All of these analysis suggest that IMLR has roughly the same
receding velocity as torus. Combined with the analysis in the
sections above, we conclude that IMLR is the inner part of a dust
torus.

\subsection{Cartoon of the Broad Line Region evolution} \label{sec-preamble}

\begin{figure}
\begin{center}
  \includegraphics[angle=0,scale=.7]{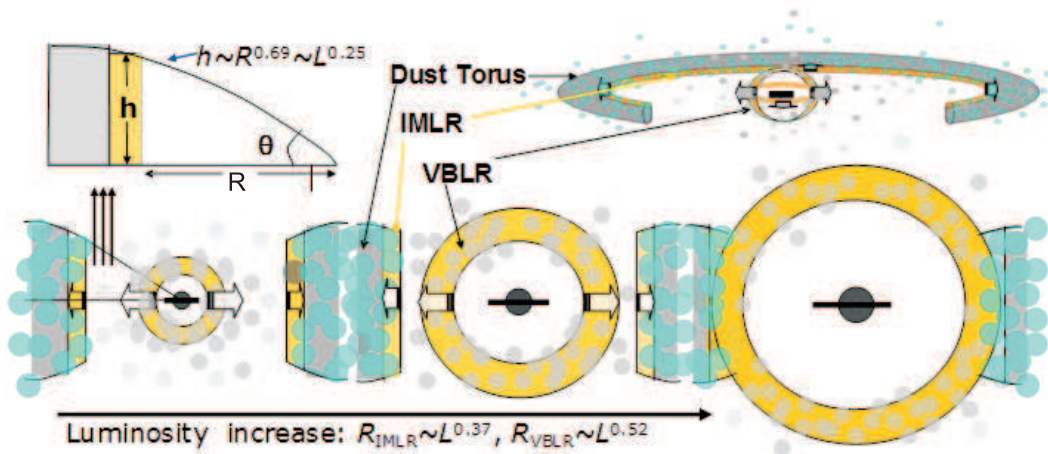}\\
  \caption{Cartoon of the Broad Line Region evolution. With increasing black hole mass and luminosity, both the
  VBLR and IMLR expand. The radius of VBLR increases faster, so the two regions have a trend to merge into
  one.}\label{fig:Car}
  \end{center}
\end{figure}

Based on the analysis above, a simple scenario of the BLR
(BLR=VBLR+IMLR) evolution can be constructed as shown in
Fig.\ref{fig:Car}. The inner spherical region is the VBLR. It
expands to a larger radius with luminosity increase and perhaps also
black hole mass increase. We suggest that the IMLR is the inner part
of the torus, which can be sublimated by the central radiation and
thus its radius also increases with luminosity increase. Naturally
the IMLR will be photoionized by the irradiation of the central AGN
and the material gravitationally bound by the central black hole
will be varialized with the gravitational potential of the central
black hole, as determined by equations (1) and (2). Since the height
of torus increases slower than the inner radius of the torus when
luminosity increases,  the covering factor of the IMLR decreases
with luminosity (and larger radius) as shown in the cartoon. When
the luminosity is high enough, the two regions may eventually merge
into one. The observed broad emission line is thus the superposition
of the emission from the two regions. When luminosity is large, the
broad line region and torus become one single entity, but with
different physical conditions. This scenario is consistent with the
dust bound BLR hypothesis (Laor 2004; Elitzur 2006). Suganuma et al.
(2006) also showed that the delay time of infrared emission is
always longer than the delay time of emission lines of the
corresponding AGN, but sometimes they are very close to each other
and become almost the same. This is also consistent with the inner
torus region as the origin of the IMGC.

\section{Other supporting evidence for the two-components model}

\subsection{The micro-lensing result of BLR in J1131-1231}

A piece of supportive evidence for the existence of IMLR comes from
the study of the micro-lensing of the Broad Line Region in the
lensed quasar J1131-1231 (Sluse et al. 2007, 2008). In this work
they found evidence that the H$\beta$ emission line (as well as
H$\alpha$) is differentially microlensed, with the broadest
component (FWHM $~$ 4000 km/s) being much more micro-lensed than the
narrower component (FWHM $~$ 2000 km/s). The emission line can be
well decomposed into two components and the emission line' profile
has been significantly changed after microlensed compared to the
original line. Because the amplitude of micro-lensing depends on the
size of the emitting region, it can be naturally explained as the
broadest component of the emission line comes from a more compact
region than that of the narrower component. Although it can also be
explained as a single region with a range of gas velocities at
different radii, the explanation comes out from our model is
natural.

\subsection{The Mrk 79' double peaks in CCCD map} \label{sec-preamble}

\begin{figure}
\begin{center}
  \includegraphics[angle=0,scale=.5]{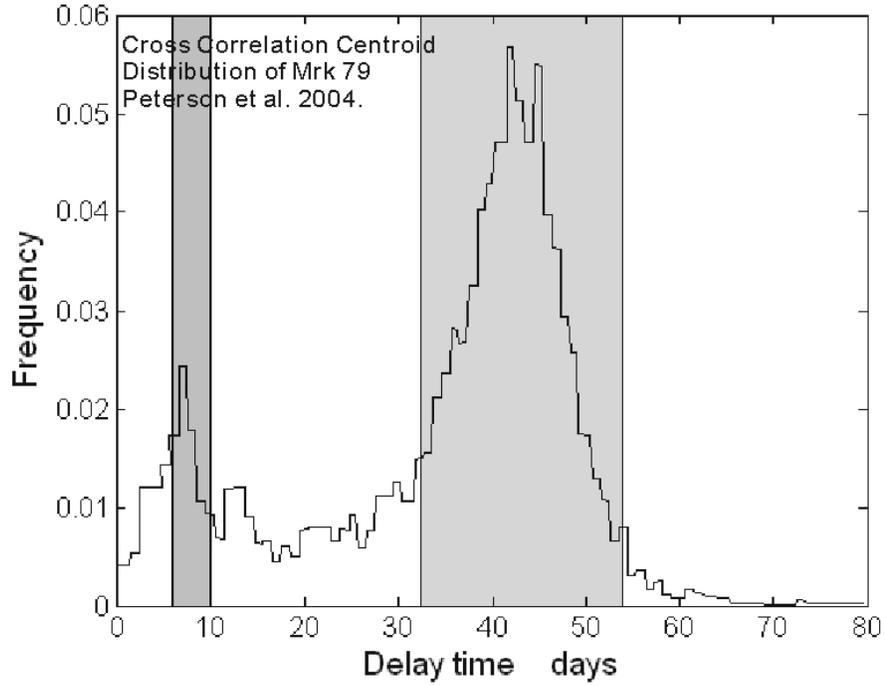}\\
  \caption{Cross Correlation Centroid Distribution of Mrk 79. The first peak is the usually used delay time
  range of broad line region which we take as the delay time of the VBLR. The second peak is the corresponding
  delay time range of the IMLR according to the FWHM ratio of their emission lines.}\label{fig:Mrk2}
  \end{center}
\end{figure}

\begin{figure}
\begin{center}
  \includegraphics[angle=0,scale=.6]{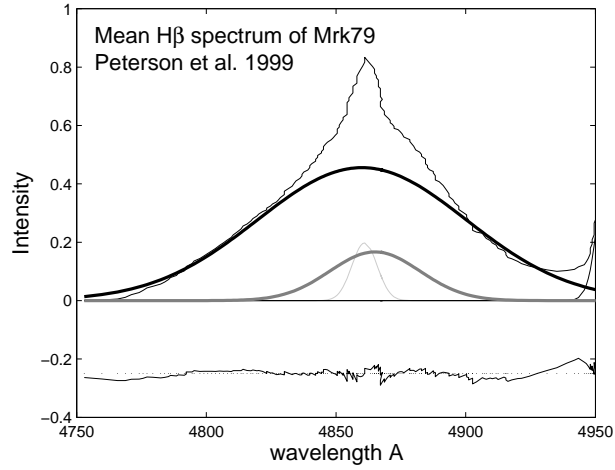}\\
  \caption{Decomposition of Mrk 79's H$\beta$ line. The green line is the normal narrow line which we do not
  discuss here. The yellow line represents the IMGC with FWHM of 2522 km$\cdot s^{-1}$, and the blue
  line represents the VBGC with FWHM of 5856 km$\cdot s^{-1}$.}\label{fig:Mrk1}
  \end{center}
\end{figure}

Then, we turn to the RM experiment. If the IMLR is really a
physically distinct emission region apart from the VBLR, another
peak corresponding to the radius of this region might appear in the
cross correlation function between the lightcurves of the emission
line and the continuum. In fact, a possible example exists in the
database of RM observations (Peterson et al. 1998; 2004). Mrk 79
shows obviously double peaks in the cross correlation centroid
distribution map as shown in Fig.\ref{fig:Mrk2}, which has not been
explained well so far. There are four subsets of RM data for Mrk 79
which were taken in different time periods, and the double-peaked
cross-correlation centroid distribution of Mrk 79 is from the fourth
period. The measured time delays are about $9-16$ days in the
subsets $1-3$ (Peterson et al. 2004), while in the 4th subset there
are two typical time delays, one is about $6-10$ days which is
consistent with subsets $1-3$, and the other is about $42$ days, as
shown in Fig.\ref{fig:Mrk2}. Here we suggest that the shorter time
delay, i.e. $6-10$ days, is the delay time of VBLR, which was also
observed in subsets $1-3$. The longer time delay, i.e. $\sim 42$
days, is possibly the delay time of IMLR, which was only observed in
the fourth subset. The mean H$\beta$ spectrum of Mrk 79 can be well
described in a three Gaussian component model (Peterson et al.
1999), i.e., a normal narrow line, an IMGC and a VBGC, as shown in
Fig.\ref{fig:Mrk1}. Based on the simple relation in section 4.1
(equation 3), we get
$R_{\rm{IMLR}}/R_{\rm{VBLR}}=V^{2}_{\rm{VBLR}}/V^{2}_{\rm{IMLR}}=(5856
 \rm{ km/s}/2522 \rm{ km/s}))^2=5.4$. Therefore the delay time for IMLR should
be 32-54 days,  fully agrees with the second peak of CCCD. This
agreement also suggests that our rather arbitrary choice of
$C_0=f_1/f_2=1$ does not deviate from its true value significantly,
at least not for Mrk 79. However, it should be noted that in the
mean H$\beta$ the VBGC dominates over the IMGC, but in the CCCD the
second peak appears to be much stronger. Probably, in the fourth
subset, the IMLR responses to the continuum more than the VBLR,
i.e., the IMGC is more variable.

On the other hand, for the majority of other sources with RM
observations, the IMGC is not important. We re-examined the ten RM
sources with public data on Peterson'
website\footnote{http://www.astronomy.ohio-state.edu/~agnwatch/}.
Six of them only need one Gaussian component to describe the broad
line of H$\beta$, one of them has double peak profile and one has
irregular profile, only two of them need VBGC+IMGC to fit. In
addition, 90\% of the sources that plot on the R-L relation (Kaspi
et al 2005) has luminosity
  larger than $10^{43}$ erg s$^{-1}$ and more than 60\% larger than $10^{44}$ erg s$^{-1}$, which prefers the
  emission line to be one single Gaussian profile (In our sample, there are 14 sources's H$\beta$
  only need one Gaussian component, 11 of them has luminosity larger than $10^{43}$ erg s$^{-1}$ and 6 of them larger than $10^{44}$ erg s$^{-1}$ ).
Therefore the Radius-Luminosity relationship established in emission
line RM measurements may be valid only for the VBLR, i.e., the R-L
relation only gives the radius of VBLR, as assumed in section 4.1.
This is probably why we did not see double peaks in most of the CCCD
maps (As shown above, even for Mrk 79, the two peaks were obtained
in only one observation period out of four periods in total.). Two
of the sources with double Gaussian line profiles are NGC 4051 and
Mrk 509; the lag times for the two components of both sources
satisfy this simple relationship
$t_{\rm{IMLR}}/t_{\rm{VBLR}}=V^{2}_{\rm{VBLR}}/V^{2}_{\rm{IMLR}}$,
consistent with our two-components BLR assumption. Interestingly the
lag time for the VBGC of NGC 4051 is between 1 to 2 days,
significantly different from the lag time of about 4 days for the
whole line. This revised lag time makes NGC 4051 consistent with the
Radius-Luminosity relationship in equation (5) or (6), resolving the
outstanding problem on the significant deviation of NGC 4051 from
the Radius-Luminosity relationship (Kaspi et al. 2005). The detailed
results on our re-analysis of the RM data with this two-components
model will be presented elsewhere (Zhu \& Zhang 2009).

\subsection{The narrower RMS spectra}
\begin{figure}
\begin{center}
  \includegraphics[angle=0,scale=.7]{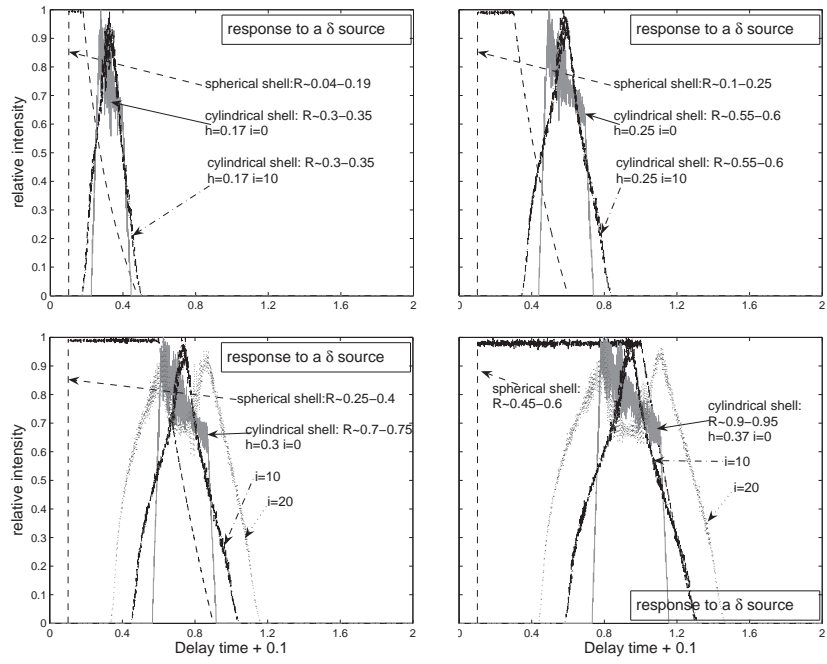}
 \caption{Different responses of different geometries to a delta impulse. Time is normalized to arbitrary unit. Each figure shows a different $R_{\rm IMLR}-R_{\rm VBLR}$ pair}\label{fig:p}
  \end{center}
\end{figure}

We propose that the VBLR and IMLR are identified with the variable regions that scales with luminosity. If this
is the case, one might expect that the VBGC would vary more than the mean spectrum as it has smaller radius.
However, that is not observed; the RMS spectrum of the emission line is normally (but not always) narrower
systematically than the mean spectrum (Collin et al. 2006). In our model, IMLR has a flattened geometry, and our
simulations show that it can create a narrower response function to a delta impulse even with radius much larger
than the VBLR. We carried out a straightforward test by calculating the responses of the two components to the
continuum. In this simple geometrical model, the VBLR is a spherical shell and the IMLR is a cylindrical shell
(representing a flattened disk-like geometry with inner and outer boundary). The gas density inside each region
is independent of radius, i.e., the gas distribution could be clustered, but the distribution of gas clusters is
independent of radius. The thickness of the VBLR shell is chosen to be three times that of the IMLR, because the
VBLR is thought to have much lower density. The emissivity law of the two emission line regions is chosen in
such a simple way that the emission line flux is proportional to the density times the continuum flux received
at any point in the two regions. This means that the broad emission line intensity is proportional to the
covering factor of each region. The line profile at any radius is assumed to have a Gaussian profile, with
velocity determined from equation (3). The overall broad emission line from each region is thus the
superposition of all broad lines from all radii. The response functions with different combinations of
parameters for the two shells and the inclination angle of IMLR are shown in Fig.\ref{fig:p}. It can be seen
that the IMLR normally can produce a narrower response function when the inclination angle is not very large. A
narrower RMS spectrum can be easily calculated based on the narrower response function of IMLR. However, the
FWHM difference in modeled RMS and mean spectra is not as significant as the observed. The simulation is carried
out with noiseless, evenly sampled data, so further work is needed to see if this model can explain the narrower
RMS spectrum quite well. In any case, the narrower RMS is not in conflict to our model of two distinct broad
line regions, with the outer region has a flattened geometry. The requirement for small inclination angles
simply suggests that most of the reverberation mapped objects have small inclination angles, a generic property
of AGNs with broad emission lines. Further more extensive calculations on the detailed responses of the two
broad line regions to characteristic continuum light curves of AGNs and comparisons with data will be presented
in a future work (Zhu \& Zhang 2009).

\section{On the two problems of AGNs}
\subsection{The under-massive black hole problem of NLS1s} \label{sec-preamble}

\begin{figure}
\begin{center}
  \includegraphics[angle=0,scale=.6]{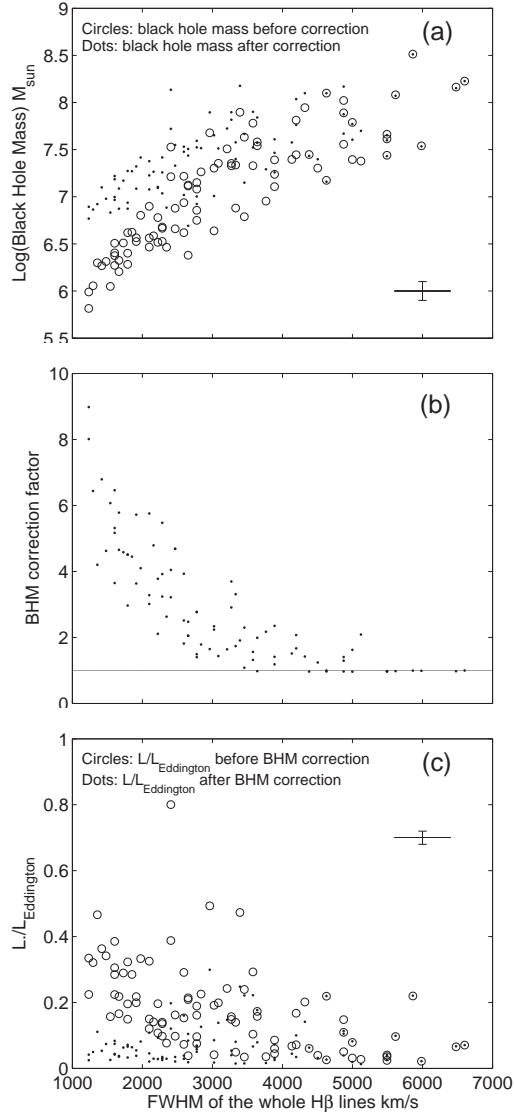}\\
   \caption{Black hole mass correction. X-axis is the FWHM of H$\beta$, y-axes are black hole
  mass, black hole mass correction
  factor and relative luminosity to $L_{\rm EDD}$ (Eddington Luminosity)
  from top to bottom. Black hole
  masses (in units of solar mass $M_{\sun}$) are plotted in logarithm scale. Circles are black hole
  mass and relative luminosity calculated with equation (1). Dots are that after correction
  using equation (4). Clearly the correction is more effective for AGNs with smaller FWHM, i.e., NLS1s. After the correction,
  their luminosity (in units of Eddington) appear to be in the same range, i.e., NLS1s do not show exceptionally higher luminosity.
  Typical uncertainty is plotted as a cross.}\label{fig:MBH}
  \end{center}
\end{figure}

RM-based black hole mass is calculated by the virial
equation\\
\begin{equation}
M_{\rm{ BH}}=\frac{R_{\rm{ BLR}}}{G}f^2\rm{FWHM}_{\rm{ H}\beta}^2,
\end{equation}
where FWHM$_{\rm{ H}\beta}$ is the FWHM of the whole H${\beta}$,
which represents the virial velocity of the BLR. For an isotropic
velocity distribution, as generally assumed, $f=\sqrt{3}/2$ (Onken
et al. 2004). $R_{\rm{ BLR}}$ is the BLR radius that can be
calculated from equation (5) or (6). As assumed in section 4.1 and
further discussed in section 5.2, the radius obtained from RM may
represent the radius of the VBLR actually. If we take this
assumption, FWHM of the VBGC, instead of FWHM of the whole line,
should be used to calculate the black hole mass. We therefore
correct the black hole mass in this way,\\
\begin{equation}
M_{\rm{ BHb}}=\frac{R_{\rm{
VBLR}}}{G}f^2\rm{FWHM}^2(\frac{\rm{FWHMb}}{\rm{FWHM}})^2,
\end{equation}
where FWHMb is the FWHM of the VBGC, and $R_{\rm{ VBLR}}$ is taken
as the $R_{\rm{ BLR}}$ in equation (5) or (6). It gives a more
significant mass correction for NLS1s than for BLS1s as shown in
Fig.\ref{fig:MBH}(a). The correction factor is near unity when FWHM
reaches about 5000 km$\cdot$s$^{-1}$. We use L5100 as an indicator
for continuum luminosity, and a correction for accretion rate has
also been shown in Fig.\ref{fig:MBH}(c). After such correction,
NLS1s still have smaller black hole masses but normal accretion rate
in units of the Eddington rate.

\begin{figure}
\begin{center}
  \includegraphics[angle=0,scale=.6]{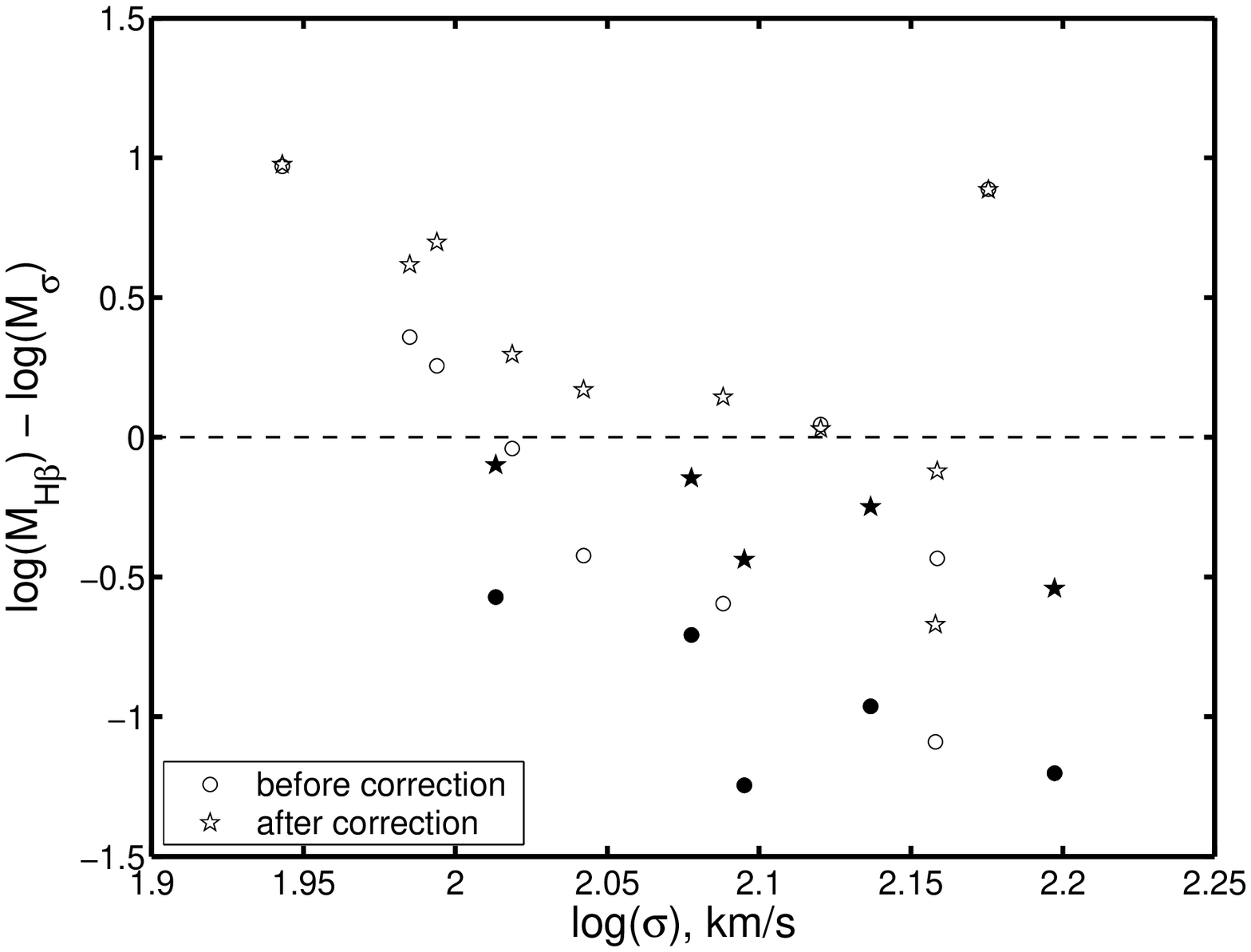}\\
 \includegraphics[angle=0,scale=.5]{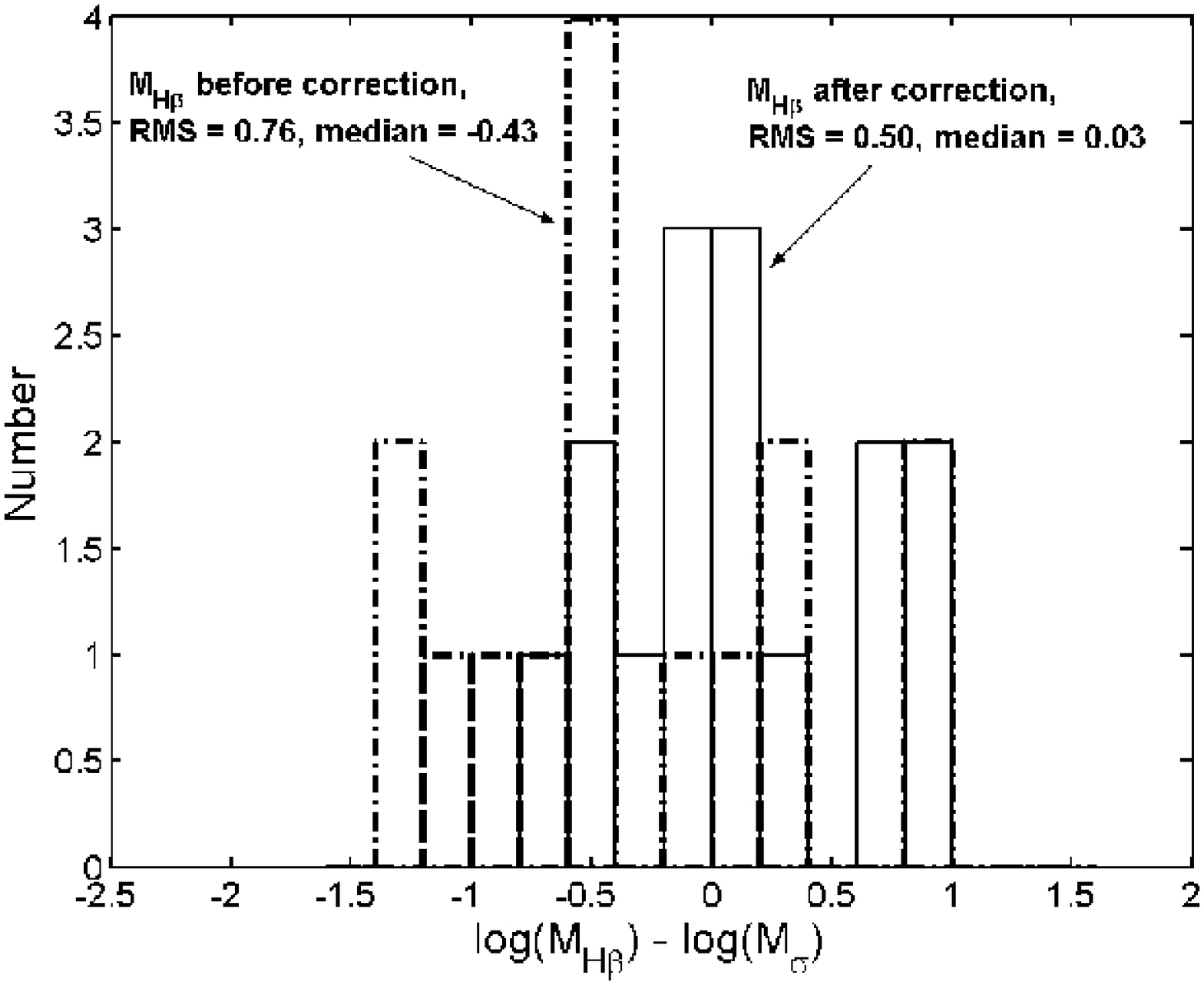}\\
  \caption{Comparison between black hole mass measured here and that predicted by the currently used M-sigma relation
  $M\propto\sigma^4$ (Tremaine et al. 2002), for fifteen sources in our sample with dispersion measurements (Shen et al. 2008). {\it Upper panel}:
   Circles represent those with black hole masses obtained with FWHM of the whole line
   used.
   Pentagrams represent those with corrected black hole mass. Filled symbols are
   NLS1s. {\it Lower panel}: Histogram of the mass ratios before and after the
   correction. After the correction, the median value of $(\log M_{{\rm H}\beta}-\log M_{\sigma})$ is much closer to zero,
   and the dispersion is reduced from 0.76 to 0.50 dex.}
   \label{fig:SM}
  \end{center}
\end{figure}

Fifteen objects in our samples have velocity dispersion measurement
data (sigma) in Shen et al. (2008), as shown in Fig.\ref{fig:SM},
where comparisons are made between black hole masses measured here
and that predicted by the currently used M-sigma relation
$M\propto\sigma^4$ (Tremaine et al. 2002). It is obvious that the
masses of all NSL1 (filled symbols in the upper panel of
Fig.\ref{fig:SM}) are well below, but become more very close to, the
predictions of the M-sigma relation before and after the correction,
respectively. For other AGNs no significant changes to their masses
are introduced by the correction process. As shown in the lower
panel of Fig.\ref{fig:SM}, after the correction, the median value of
$(\log M_{{\rm H}\beta}-\log M_{\sigma})$ is much closer to zero,
and the dispersion is reduced from 0.76 to 0.50 dex, which is
consistent with the black hole mass uncertainty of 0.5 dex in RM
(Peterson 2006). The effect of the correction is obvious, albeit
small number statistics due to the limited sample.

\subsection{The under-populated luminous type II problem of Seyfert galaxies}

It has been shown that the receding torus model with constant torus
height fails to provide a good fit to the data of type II AGN
fraction as a function luminosity, and a good fitting can be given
when $h$ increases slowly with luminosity with the relation
$h\propto L^{0.23}$ (Simpson 2005), in excellent agreement with that
of IMLR as we have shown in section 4.4. Note that here we consider
the change of covering factor is totally because of the change of
the opening angle during the calculation, following Wang et al.
(2005). This agreement suggests that the receding of torus is
sufficient to explain the decrease of covering factor with
decreasing luminosity. Our model is consistent with the torus that
Simpson (2005) needed to explain under-populated luminous type II
AGNs.

\section{Conclusion and discussion}

We conclude that the decomposition of broad H$\beta$, H$\alpha$, and
H$\gamma$ line of the AGNs confirms the two component model of BLR
which has been suggested by several previous studies (e.g.,
Brotherton 1996; Sulentic et al. 2000; Hu et al. 2008.). We have
made detailed analysis about the two emission regions (VBLR and
IMLR) based on the Balmer line decomposition, and find other
supportive evidence for this model. Our main conclusions are:

\begin{enumerate}

\item The two Gaussian components exhibit evolutions with increasing
FWHM
 (we note in passing that because the two components show
quite different dependence with FWHM and luminosity as to be shown
in the following, we rule out the possibility that the dependence is
caused by systematic biases in the decomposition process). The
evolution is much stronger for the H$\beta$ and H$\gamma$ lines. Our
results offer strong evidence of the evolution of the broad line
region (consists of a IMLR and a VBLR) evolution from NLS1s to
BLS1s. We obtain the luminosity dependence for the radius of IMLR,
$R_{\rm IMLR}\propto L^{0.37}$, if the luminosity dependence for the
VBLR is taken as $R_{\rm VBLR}\propto L^{0.52}$ (Kaspi et al. 2005).
The two emission regions have a trend to merge into one region with
luminosity increasing. Balmer decrement and the Baldwin effect in
IMLR indicate that it has a flattened geometry, higher density and
contains more dust, compared to the VBLR. The receding velocity of
IMLR is consistent with dust torus. Therefore, we suggest the IMLR
is the hot inner skin of torus. A cartoon of the evolution of BLR
emerge from these analysis.

\item There are other evidence in support of this two component
model. The study of micro-lensing provides possible evidence for the
existence of IMLR. The double-peaked CCCD from the RM data of Mrk 79
also provides possible evidence. Simulations suggest that the
narrower RMS spectra of broad emission lines from many AGNs may be
consistent with our model, although more work needs to be done to
establish this as the case. The existence of a weak IMGC in the
broad emission lines of many sources with RM measurements may cause
systematic biases for the measurements of the radius of the VBLR,
thus biasing the Radius-Luminosity relation for the VBLR. It will be
helpful to decompose each broad emission line into two Gaussian
components as we have done here, and then do cross-correlation
analysis between the continuum and each of the two components, in
order to measure the Radius-Luminosity relations for the two
components independently.

\item In our model, only the VBGC should be used to estimate the
black hole mass, and the radius measured by reverberation mapping
based on emission lines normally represents the radius of this
region. After correction for black hole masses, NLS1s still have
smaller black hole masses (compared to BLS1s) but normal accretion
rate in units of Eddington rate. Therefore, the black hole mass
increases from NLS1s to BLS1s by following the M-sigma relation
established for normal galaxies.

We obtain the luminosity dependence for the height of IMLR as
$h\propto L^{0.25}$. It can well explain the luminosity function of
AGN, if the decreasing fraction of type II AGNs for higher
luminosity is due to completely the decreasing covering angle of the
dusty torus to the central irradiation source (Simpson. 2005).

Therefore both the problem of under-massive black hole in NLS1s and
the problem of under-populated luminous type II Seyfert galaxies can
be understood properly if our model is true, still more concrete
evidences for this model are needed.

\end{enumerate}

 Hu et al. (2008) found evidence that IMLR is related to inflow
towards the VBLR. Combining with this conclusion, we suggest that
the inflow from the inner boundary region of AGN's dusty torus may
provide the supply to the accretion disk surrounding the central
black hole. The strong and positive luminosity dependence of the
geometry of IMLR suggests that the dust sublimation by the central
accreting black hole's radiation dominates the structure and
evolution of IMLR. If IMLR is related to inflow (Hu et al. 2008), we
may be able to further suggest that the inflow is caused by the dust
sublimation, i.e., a consequence of the feed-back of the black
hole's accretion and radiation. Because IMLR is also ionized
similarly to VBLR, the IMLR induced viscosity allows efficient
angular momentum transfer to drive gas inflow from the dust torus
and consequently fuel the accretion flow onto the central black
hole. In this scenario, the accretion flow is self-regulated by the
radiation from the accretion disk through irradiation to the dust
torus. Therefore the growth of the supermassive black hole is at the
expense of consuming the material in the dust torus during the AGN
phase; this is consistent with the observation that for very low
luminosity AGNs, the luminosity decreases with decreasing absorption
column, i.e., the AGNs in their last stages are running out
accretion material supplied by the torus (Zhang \& Soria et al.
2009). Of course not all material in the accretion flow falls into
the black hole horizon to increase the black hole'a mass, since
accretion winds and outflows are common in AGNs.

\begin{figure}
\begin{center}
  \includegraphics[angle=0,scale=.6]{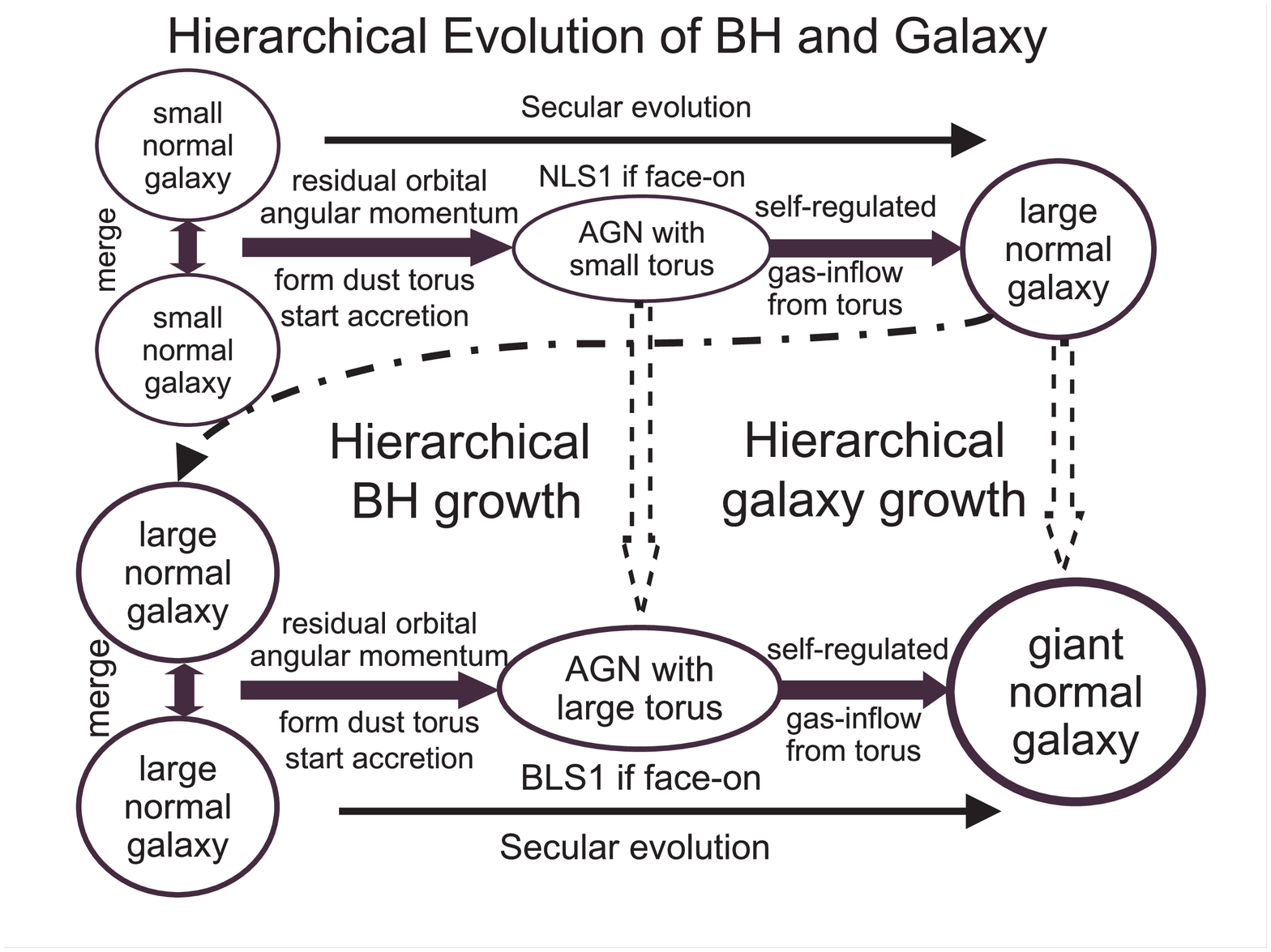}\\
  \caption{Illustration of hierarchical evolution of black hole, torus and host galaxy. Two galaxies merge to
  form a dust torus, due to the residual orbital angular momentum of the two galaxies in a binary. The dust torus is
  sublimated by the irradiation of the accretion disk around the central black hole. The gas produced by the
  sublimation process is ionized
  also by the irradiation of the accretion disk around the central black hole. The MRI viscosity in the ionized gas transfers
  the angular momentum outwards and thus fuels the accretion flow onto the central black hole;
  this is the AGN phase of a galaxy. Such self-regulated process
  grows the black hole by consuming the dust material in the torus,
  until the torus disappears and the AGN activity is turned off, i.e.,
  a normal, inactive and larger galaxy is formed. In this scenario, NLS1 galaxies are the AGN phase (viewed through
  the opening cone of the dust torus) resulted from the merger
  of two smaller (dwarf) galaxies. Subsequent mergers of larger galaxies progressively
  produce AGNs with larger tori, higher luminosity and more massive black holes, in a
  hierarchical evolutionary sequence. It should be noted that here we only illustrate the processes of major mergers, in order
  to emphasize our main points, despite that fact that minor mergers are more frequent.}\label{fig:evolution}
 \end{center}
\end{figure}

After the material in the dust torus is completely consumed, the AGN
phase will be turned off and the galaxy becomes a normal and
inactive galaxy. Indeed many AGNs in low luminosity (because of low
accretion rate and low radiation efficiency) show very little, or
even no signs of torus and/or broad line region; there is also no
evidence for dust torus or broad line region in the centers of
normal and inactive galaxies, including the Milky Way. Therefore the
AGN's dust torus is the missing link or bridge between the coeval
growth of a black hole and its host galaxy. This would require that
during the merging of two galaxies, a dust torus is first formed,
perhaps due to the residual orbital angular momentum of the two
galaxies. The dust torus then fuels the accretion and growth of the
supermassive black hole through the self-regulation of irradiation
to the dust torus by the accretion disk. The initial trigger to this
self-regulation process may be Bondi accretion of gas with
negligible angular momentum, or the low level AGN activities of the
two black holes in the two parent galaxies before the merger. This
evolutionary scenario is illustrated in Fig.\ref{fig:evolution}.

The above scenario is generally consistent with that proposed by
Wang \& Zhang (2007),  but also with some important difference. We
suggest that the torus is formed by the merging of two galaxies and
disappears after each AGN cycle; the different appearance (mainly
geometry) of torus in different types of AGNs are mostly due to the
self-regulation of the accretion and irradiation of the AGN.
Therefore in our scenario, torus evolution is fast (only lasting for
one episode of AGN activity) and synchronized with that of the broad
line region, whose evolution is also dominated by the luminosity of
the AGN. {\it Our torus evolution is mostly hierarchical.} For
example, although NLS1s also have generally smaller black hole
masses, we do not find NLS1s deviate from the M-sigma relation more
than the BLS1s after black hole mass correction made here. Therefore
in our scenario NLS1s are produced by mergers of smaller galaxies
compared to BLS1s; NLS1 may or may not show up as BLS1s in the
future, depending upon if more galaxy mergers grow them up in the
future. In the scenario of Wang \& Zhang (2005), NLS1s are in their
early growth stage and will grow to become BLS1 during this
particular AGN cycle. {\it Therefore their torus evolution is mostly
secular.} In practice, both hierarchical and secular evolutions
should be needed for the black hole, torus and host galaxy. It is
natural that hierarchical evolution dominates at high redshifts
where merger rate is very high, and secular evolution dominates at
low redshifts.

\acknowledgments

We are extremely grateful to G. La Mura and L.C. Popovic for sending
us the processed spectra with both the narrow lines and continuum
already removed, for the sample we used in this work. SNZ thanks
Jianmin Wang for many discussions, as well as sharing the early
results of their work on the similar subject prior to its
publication (Hu et al. 2008), which motivated us to pursue the work
presented here. Xuebing Wu, Chen Hu and Yuan Liu are thanked for
their comments and suggestions, when the initial results of this
work was presented in a black hole workshop organized by Feng Yuan,
Xinwu Cao and Wenfei Yu during April 26-28, 2008, in Shanghai
Astronomical Observatory. Yuan Liu, Jianmin Wang, Chen Hu and Xinlin
Zhou are also appreciated for proof reading and suggestions on the
draft of this manuscript. Zhixing Ling is acknowledged for helping us in calculating
the responses functions presented in section 5.6 (Fig. 17). We appreciate very much the insightful
comments and help suggestions by anonymous referee. SNZ acknowledges
partial funding support by the Yangtze Endowment from the Ministry
of Education at Tsinghua University, Directional Research Project of
the Chinese Academy of Sciences under project No. KJCX2-YW-T03 and
by the National Natural Science Foundation of China under grant Nos.
10521001, 10733010,10725313, and by 973
Program of China under grant 2009CB824800.

%% Appendix material should be preceded with a single \appendix command.
%% There should be a \section command for each appendix. Mark appendix
%% subsections with the same markup you use in the main body of the paper.

%% Each Appendix (indicated with \section) will be lettered A, B, C, etc.
%% The equation counter will reset when it encounters the \appendix
%% command and will number appendix equations (A1), (A2), etc.

\begin{deluxetable}{|lllll|lllll|lllll|lllll|}
\tabletypesize{\scriptsize}
\rotate
\tablecolumns{20}
\tablewidth{0pt}
\setlength\tabcolsep{5pt}
 \tablecaption{Source properties and decomposition parameters. L5100 are given in units of $10^{42}$ erg $\rm{s}^{-1}$, ${M}_{\rm{BH}}$ are expressed in $10^5$ ${M}_{\sun}$. (1): Object name; (2): redshift; (3): L5100; (4): $M_{\rm{BH}}$; (5): Hb (height of the VBGC); (6): Hi (height of IMGC); (7): FWHM; (8): FWHMb; (9): FWHMi; (5), (6), (7), (8), (9) are parameters of H$\alpha$ lines; (10), (11), (12), (13), (14) are the corresponding parameters of H$\beta$ lines; (15), (16), (17), (18), (19) are the corresponding parameters of H$\gamma$ lines.}
\tablehead{ \colhead{} & \colhead{} & \colhead{} & \colhead{} &
\colhead{} & \multicolumn{5}{c}{H$\alpha$ lines}&
\multicolumn{5}{c}{H$\beta$ lines}&
\multicolumn{5}{c}{H$\gamma$ lines}\\
\colhead{NO} & \colhead{(1)}  &
\colhead{(2)} & \colhead{(3)} & \colhead{(4)} &
\colhead{(5)} & \colhead{(6)} & \colhead{(7)} &
\colhead{(8)} & \colhead{(9)} & \colhead{(10)} &
\colhead{(11)} & \colhead{(12)} & \colhead{(13)} &
\colhead{(14)} & \colhead{(15)} & \colhead{(16)} &
\colhead{(17)} & \colhead{(18)} & \colhead{(19)} }

\startdata
  1  & SDSSJ1152-0005 &  0.275  &  77.88  &  208.85  &  18.9  &  0     &  3610  &  3579  &  0     &  13.6   &  0    &  3640  &  3601  &  0     &  4.26   &  2.15   &  2529  &  3913  &  1493\\
  2  & SDSSJ1157-0022 &  0.178  &  140.86 &  1241.78 &  75.2  &  58.5  &  4844  &  6926  &  3241  &  48.56  &  0    &  6602  &  6598  &  0     &  15.7   &  5.77   &  4689  &  6627  &  2301\\
  3  & SDSSJ1307-0036 &  0.188  &  48.84  &  149.05  &  26.9  &  36.9  &  1873  &  3770  &  1285  &  11.3   &  8.45 &  2838  &  3793  &  1857  &  7.24   &  2.76   &  3023  &  4143  &  1597\\
  4  & SDSSJ1059-0005 &  0.283  &  92.33  &  224.16  &  16.2  &  24.4  &  1965  &  3839  &  1368  &  8.17   &  5.66 &  3208  &  4109  &  2224  &  5.8    &  0      &  4010  &  4362  &  0\\
  5  & SDSSJ1342-0053 &  0.129  &  23.61  &  293.32  &  14.3  &  27.8  &  2925  &  5767  &  2272  &  7.43   &  6    &  4195  &  6041  &  2716  &  3.43   &  3.39   &  3763  &  6445  &  2840\\
  6  & SDSSJ1307+0107 &  0.26   &  183.28 &  1056.59 &  40.6  &  28.1  &  4433  &  6240  &  2864  &  16.4   &  8.38 &  4874  &  5784  &  3311  &  11.4   &  0      &  4380  &  4851  &  0\\
  7  & SDSSJ1341-0053 &  0.17   &  32.02  &  97.72   &  22.9  &  57.5  &  2285  &  4490  &  1835  &  12.4   &  14.1 &  2653  &  4173  &  1869  &  6.15   &  8.04   &  2406  &  4791  &  1853\\
  8  & SDSSJ1344+0005 &  0.276  &  113    &  1133.83 &  31.1  &  24.9  &  5027  &  6849  &  3541  &  18.8   &  0    &  6478  &  6396  &  0     &  7.38   &  0      &  4751  &  5269  &  0\\
  9  & SDSSJ1013-0052 &  0.327  &  324.69 &  2732.11 &  38.1  &  0     &  4342  &  4257  &  0     &  23.1   &  0    &  4627  &  4623  &  0     &  11.1   &  0      &  4504  &  4937  &  0\\
  10 & SDSSJ1010+0043 &  0.237  &  ...    &  ...     &  ...   &  ...   &  ...   &  ...   &  ...   &  ...    &  ...  &  ...   &  ...   &  ...   &  ...    &  ...    &  ...   &  ...   &  ...\\
  11 & SDSSJ1057-0041 &  0.087  &  5.28   &  42.13   &  7.78  &  51.8  &  2148  &  4946  &  1921  &  8.43   &  17.9 &  2468  &  5342  &  1894  &  4.84   &  10.3   &  2406  &  4028  &  2211\\
  12 & SDSSJ0117+0000 &  0.245  &  33.58  &  124.96  &  27.5  &  34.6  &  1828  &  3547  &  1185  &  13.1   &  8.99 &  2653  &  3801  &  1533  &  5.59   &  4.3    &  2591  &  4248  &  1853\\
  13 & SDSSJ0112+0003 &  0.074  &  4.16   &  32.75   &  15.4  &  41    &  1873  &  4566  &  1487  &  7      &  10.2 &  2221  &  4318  &  1527  &  0      &  9.48   &  1789  &  0     &  1923\\
  14 & SDSSJ1344-0015 &  0.141  &  30.36  &  78.27   &  24.5  &  68.4  &  1919  &  3881  &  1550  &  12.9   &  23.6 &  2097  &  3643  &  1598  &  7.85   &  10     &  2344  &  4208  &  1826\\
  15 & SDSSJ1343+0004 &  0.114  &  25.02  &  57.24   &  27.5  &  71    &  1736  &  3639  &  1391  &  10.9   &  25.9 &  1974  &  3997  &  1559  &  7.58   &  10.8   &  1974  &  3059  &  1633\\
  16 & SDSSJ1519+0016 &  0.233  &  100.02 &  628.19  &  35.6  &  16.8  &  3884  &  5251  &  2109  &  16.1   &  0    &  4874  &  4793  &  0     &  7.32   &  0      &  4010  &  4481  &  0\\
  17 & SDSSJ1437+0007 &  0.179  &  74.76  &  146.48  &  21    &  63.6  &  1965  &  4012  &  1631  &  11.7   &  22.7 &  2406  &  4315  &  1877  &  8.27   &  2.21   &  3578  &  4377  &  2278\\
  18 & SDSSJ1659+6202 &  0.31   &  210.04 &  752.96  &  16.9  &  36.1  &  3564  &  5479  &  2955  &  12.9   &  8.59 &  4319  &  5154  &  3303  &  4.88   &  10.9   &  2776  &  3989  &  2655\\
  19 & SDSSJ0121-0102 &  0.36   &  439.86 &  724.5   &  18.5  &  50.4  &  2833  &  4770  &  2377  &  14.6   &  15.4 &  3393  &  4688  &  2599  &  26.4   &  0      &  4134  &  4506  &  0\\
  20 & SDSSJ1719+5937 &  0.174  &  137.49 &  946.25  &  117   &  60.5  &  4936  &  6088  &  3232  &  56.2   &  0    &  5615  &  5550  &  0     &  12.2   &  0      &  3640  &  4077  &  0\\
  21 & SDSSJ1717+5815 &  0.279  &  208.76 &  511.13  &  38.6  &  25.6  &  2925  &  3814  &  1902  &  14.4   &  14.5 &  3578  &  4109  &  3076  &  7.64   &  0      &  3023  &  3381  &  0\\
  22 & SDSSJ0037+0008 &  0.362  &  277.83 &  369.99  &  33.2  &  55.2  &  1416  &  3424  &  948   &  14.7   &  6.71 &  2961  &  3801  &  1724  &  3.2    &  1.48   &  3332  &  4224  &  1708\\
  23 & SDSSJ2351-0109 &  0.252  &  64.75  &  246.21  &  10    &  11.8  &  3107  &  4733  &  2283  &  5.98   &  3.64 &  3640  &  5137  &  1907  &  0      &  18.6   &  2159  &  0     &  2382\\
  24 & SDSSJ2349-0036 &  0.046  &  2.13   &  35.98   &  27.2  &  72    &  2468  &  4222  &  2053  &  16.7   &  20.6 &  3023  &  4623  &  2236  &  4.84   &  5.69   &  3702  &  6560  &  2929\\
  25 & SDSSJ0013+0052 &  0.239  &  128.04 &  639.96  &  16.4  &  34.8  &  3427  &  5892  &  2795  &  12.3   &  7.01 &  4195  &  5423  &  2505  &  13.7   &  17     &  2838  &  4258  &  2418\\
  26 & SDSSJ1720+5540 &  0.055  &  19.96  &  175.4   &  64.5  &  78.7  &  4021  &  5098  &  3296  &  39.8   &  22.5 &  4134  &  5085  &  2790  &  7.96   &  3.14   &  2838  &  3607  &  1697\\
  27 & SDSSJ0256+0113 &  0.081  &  6.76   &  104.17  &  26.2  &  22    &  3336  &  4251  &  2474  &  13.9   &  4.93 &  3887  &  4623  &  2188  &  23.2   &  0      &  5121  &  5676  &  0\\
  28 & SDSSJ0135-0044 &  0.335  &  845.06 &  2960.01 &  101   &  19    &  4981  &  5451  &  2588  &  46.6   &  0    &  5861  &  5850  &  0     &  5.63   &  5.91   &  2714  &  4023  &  1963\\
  29 & SDSSJ0140-0050 &  0.146  &  22.95  &  107.38  &  23.9  &  38.7  &  2148  &  4241  &  1577  &  13.5   &  13.6 &  2776  &  4623  &  1499  &  9.19   &  0      &  4380  &  4792  &  0\\
  30 & SDSSJ0310-0049 &  0.206  &  57.01  &  471.05  &  24.9  &  45.7  &  3245  &  5174  &  2617  &  23.7   &  0    &  4997  &  4900  &  0     &  3.73   &  3.04   &  2776  &  3913  &  2123\\
  31 & SDSSJ0304+0028 &  0.368  &  120.96 &  306.39  &  19.6  &  0     &  3153  &  3066  &  0     &  8.99   &  1.83 &  3455  &  3595  &  1980  &  4.37   &  4.78   &  3023  &  5507  &2346\\
  32 & SDSSJ0159+0105 &  0.198  &  37.36  &  188.85  & 12.4   &  41.2  &  2605  &  6198  &  2179  &  7.88   &  11.4 &  3270  &  6288  &  2305  &  7.99   &  0      &  3270  &  3568  &  0\\
  33 & SDSSJ0233-0107 &  0.177  &  26.13  &  184.98  &  22.5  &  40.8  &  2605  &  4514  &  2046  &  11.3   &  7.13 &  3578  &  4472  &  2335  &  ...    &  ...    &  ...   &  ...   &  ...\\
  34 & SDSSJ0250+0025 &  0.045  &  2.07   &   9.93   &  18.6  &  94.1  &  1188  &  3439  &  981   &  9.84   &  28.2 &  1542  &  3801  &  1214  &  0      &  16.6   &  1542  &  0     &  1634\\
  35 & SDSSJ0409-0429 &  0.081  &  41.92  &  213.82  &  54.9  &  203   &  2650  &  5935  &  2254  &  28.9   &  58.7 &  3270  &  5576  &  2588  &  17.3   &  17.8   &  3270  &  5870  &  2463\\
  36 & SDSSJ0937+0105 &  0.108  &  14.45  &  63.25   &  20.7  &  98.4  &  2102  &  4946  &  1813  &  11.7   &  32.8 &  2468  &  5342  &  1969  &  5.64   &  17.8   &  2406  &  4028  &  2345\\
  37 & SDSSJ0323+0035 &  0.186  &  318.83 &  323.42  &  109   &  216   &  2239  &  4416  &  1710  &  54.9   &  84.1 &  2406  &  4841  &  1725  &  37.1   &  21.5   &  2714  &  4016  &  1617\\
  38 & SDSSJ0107+1408 &  0.216  &  56.76  &  120.52  &  28.6  &  54.7  &  1416  &  3194  &  1043  &  13.9   &  9.74 &  2591  &  3493  &  1699  &  4.76   &  6.17   &  2468  &  4678  &  1944\\
  39 & SDSSJ0142+0005 &  0.077  &  1.73   &  5.59    &  6.31  &  34.7  &  1051  &  3424  &  882   &  4.19   &  13.2 &  1234  &  3698  &  962   &  2.41   &  5.99   &  1419  &  3683  &  1251\\
  40 & SDSSJ0306+0003 &  0.095  &  3.14   &  12.56   &  14.1  &  39.2  &  1096  &  3262  &  844   &  5.99   &  8.83 &  1666  &  3595  &  1097  &  1.79   &  3.88   &  2159  &  4028  &  1897\\
  41 & SDSSJ0322+0055 &  0.09   &  6.44   &  33.31   &  18.9  &  56    &  1142  &  3442  &  917   &  4.86   &  10.9 &  2097  &  3801  &  1699  &  ...    &  ...    &  ...   &  ...   &  ...\\
  42 & SDSSJ0150+1323 &  0.037  &  8.78   &  227.13  &  127   &  76.3  &  3473  &  5450  &  1703  &  45.07  &  0    &  5985  &  5949  &  0     &  19.7   &  0      &  4010  &  4488  &  0\\
  43 & SDSSJ0855+5252 &  0.069  &  5.81   &  23.72   &  31.7  &  109   &  1051  &  2739  &  834   &  12.5   &  20.7 &  1789  &  3082  &  1363  &  0      &  15.1   &  1542  &  0     &  1591\\
  44 & SDSSJ0904+5536 &  0.039  &  1.09   &  20.98   &  20.9  &  33.8  &  1828  &  3354  &  1389  &  9.34   &  7.07 &  2653  &  3801  &  1646  &  ...    &  ...    &  ...   &  ...   &  ...\\
  45 & SDSSJ1355+6440 &  0.051  &  9.14   &  23      &  110   &  225   &  1325  &  2843  &  1010  &  54.6   &  87.1 &  1604  &  3063  &  1157  &  25.2   &  36     &  1666  &  2720  &  1405\\
  46 & SDSSJ0351-0526 &  0.075  &  31.62  &  120.58  &  83.7  &  255   &  2422  &  4566  &  2022  &  54.3   &  113  &  2776  &  4620  &  2183  &  0      &  76.9   &  2529  &  0     &  2831\\
  47 & SDSSJ1505+0342 &  0.058  &  7.98   &  18.08   &  23.2  &  98.9  &  1416  &  3887  &  1233  &  12.4   &  32.6 &  1604  &  4077  &  1254  &  8.47   &  14.3   &  2036  &  3309  &  1727\\
  48 & SDSSJ1203+0229 &  0.093  &  35.99  &  171.89  &  56    &  124   &  2559  &  4718  &  2051  &  41.9   &  27.2 &  3332  &  4392  &  2216  &  10.6   &  21.6   &  2961  &  5409  &  2704\\
  49 & SDSSJ1246+0222 &  0.078  &  45.44  &  170.3   &  41.3  &  148   &  2650  &  4870  &  2330  &  30.2   &  46.4 &  3023  &  4520  &  2348  &  16.4   &  27.3   &  2653  &  3875  &  2468\\
  50 & SDSSJ0839+4847 &  0.024  &  7.86   &  279.34  &  289   &  113   &  3976  &  5228  &  1568  &  130    &  0    &  5491  &  5428  &  0     &  63.9   &  0      &  4874  &  5327  &  0\\
  51 & SDSSJ0925+5335 &  0.087  &  5.44   &  17.86   &  7.9   &  36.8  &  1234  &  3757  &  1032  &  4.04   &  10.1 &  1666  &  4006  &  1272  &  ...    &  ...    &  ...   &  ...   &  ...\\
  52 & SDSSJ1331+0131 &  0.048  &  11.05  &  26.69   &  41.9  &  172   &  1188  &  3294  &  987   &  25.6   &  61.7 &  1727  &  3698  &  1338  &  9.19   &  34.3   &  1480  &  4143  &  1312\\
  53 & SDSSJ1042+0414 &  0.08   &  7.41   &  38.49   &  7.82  &  51.3  &  2193  &  4946  &  2013  &  6.93   &  13.3 &  2283  &  4520  &  1734  &  0      &  11.8   &  2283  &  0     &  2428\\
  54 & SDSSJ1349+0204 &  0.033  &  2.48   &  45.98   &  111   &  220   &  2468  &  4908  &  1920  &  42.9   &  27   &  3455  &  5239  &  1558  &  ...    &  ...    &  ...   &  ...   &  ...\\
  55 & SDSSJ1223+0240 &  0.072  &  8.15   &  59.21   &  35.8  &  27.9  &  1965  &  3044  &  1155  &  15     &  6.49 &  2776  &  3287  &  1777  &  11.5   &  0      &  2653  &  2959  &  0\\
  56 & SDSSJ0755+3911 &  0.034  &  7.88   &  26.39   &  58.2  &  164   &  1462  &  3321  &  1172  &  34.3   &  51.2 &  1912  &  3647  &  1344  &  ...    &  ...    &  ...   &  ...   &  ...\\
  57 & SDSSJ1141+0241 &  0.047  &  3.37   &  26.76   &  27.4  &  53    &  1599  &  3576  &  1159  &   8.6   &  13.1 &  1789  &  3801  &  1221  &  ...    &  ...    &  ...   &  ...   &  ...\\
  58 & SDSSJ1122+0117 &  0.04   &  4.94   &  17.08   &  45.6  &  144   &  1371  &  3298  &  1079  &  17     &  40.1 &  1604  &  3698  &  1235  &  ...    &  ...    &  ...   &  ...   &  ...\\
  59 & SDSSJ1243+0252 &  0.077  &  4.31   &  9.24    &  13.8  &  84.3  &  1096  &  3292  &  961   &  6.56   &  29.6 &  1295  &  3287  &  1035  &  0      &  15.2   &  1295  &  0     &  1383\\
  60 & SDSSJ0832+4614 &  0.061  &  7.87   &  44.3    &  47.2  &  152   &  1965  &  5297  &  1624  &  24.3   &  44.9 &  2283  &  5342  &  1657  &  6.99   &  24.1   &  2776  &  6312  &  2588\\
  61 & SDSSJ0840+0333 &  0.053  &  4.57   &  124.7   &  40.9  &  0     &  3884  &  3805  &  0     &  15.5   &  0    &  4627  &  4551  &  0     &  10.2   &  0      &  2653  &  2920  &  0\\
  62 & SDSSJ1510+0058 &  0.036  &  21.63  &  365.27  &  212   &  276   &  3702  &  5935  &  2735  &  127    &  25.1 &  4874  &  5548  &  1735  &  31     &  48.1   &  3085  &  6675  &  2506\\
  63 & SDSSJ0110-1008 &  0.078  &  53.19  &  187.64  &  234   &  164   &  2102  &  3372  &  997   &  99.1   &  38.8 &  3085  &  3698  &  1376  &  59     &  0      &  3023  &  3288  &  0\\
  64 & SDSSJ0142-1008 &  0.031  &  20.11  &  266.31  &  416   &  256   &  3884  &  4946  &  2631  &  238    &  0    &  4380  &  4303  &  0     &  103    &  31.3   &  4134  &  5524  &  2405\\
  65 & SDSSJ1519+5908 &  0.069  &  7.95   &  15.03   &  25.8  &  122   &  1279  &  3337  &  1086  &  14     &  51.4 &  1419  &  3698  &  1157  &  0      &  26.1   &  1542  &  0     &  1681\\
  66 & SDSSJ0013-0951 &  0.074  &  3.86   &  8.08    &  9.59  &  53.5  &  1051  &  3216  &  899   &  4.61   &  18   &  1234  &  3493  &  1027  &  ...    &  ...    &  ...   &  ...   &  ...\\
  67 & SDSSJ1535+5754 &  0.062  &  12.3   &  168.21  &  18.9  &  60.5  &  2970  &  5479  &  2569  &  14.3   &  17.4 &  3887  &  5959  &  2823  &  12.7   &  1.59   &  3085  &  3470  &  1732\\
  68 & SDSSJ1654+3925 &  0.042  &  3.6    &  35.71   &  36    &  155   &  2148  &  4490  &  1834  &  19.7   &  33.2 &  2591  &  4109  &  2054  &  0      &  23.2   &  2406  &  0     &  2562\\
  69 & SDSSJ0042-1049 &  0.058  &  3.89   &  33.23   &  20.4  &  69.7  &  1599  &  3757  &  1318  &  6.31   &  11.9 &  2283  &  4109  &  1729  &  6.51   &  2.04   &  2406  &  3377  &  650\\
  70 & SDSSJ2058-0650 &  0.09   &  4.15   &  28.15   &  13.6  &  51.4  &  1736  &  4598  &  1486  &  8.96   &  14   &  2097  &  5034  &  1420  &  0      &  23.7   &  3393  &  0     &  3752\\
  71 & SDSSJ1300+6139 &  0.052  &  9.25   &  214.11  &  67    &  49.3  &  3656  &  4948  &  2398  &  31.7   &  9.2  &  4504  &  5021  &  2657  &  1.43   &  4.69   &  1727  &  4834  &  1626\\
  72 & SDSSJ0752+2617 &  0.095  &  6.39   &  33.3    &  7.76  &  24.8  &  1599  &  4367  &  1283  &  4.35   &  7.55 &  2159  &  4726  &  1549  &  5.18   &  28     &  1480  &  4143  &  1430\\
  73 & SDSSJ1157+0412 &  0.082  &  14.61  &  28.9    &  53.9  &  163   &  1234  &  3325  &  982   &  25     &  55.9 &  1604  &  3647  &  1246  &  6.82   &  15.3   &  2221  &  3913  &  2064\\
  74 & SDSSJ1139+5911 &  0.085  &  15.6   &  77.92   &  17.9  &  58.8  &  2330  &  4718  &  1969  &  9.91   &  26.1 &  2591  &  5137  &  2110  &  ...    &  ...    &  ...   &  ...   &  ...\\
  75 & SDSSJ1345-0259 &  0.028  &  4.39   &  65.57   &  18.4  &  144   &  2742  &  5859  &  2515  &  26.8   &  44.6 &  3332  &  6061  &  2521  &  0      &  38.6   &  2653  &  0     &  2969\\
  76 & SDSSJ1118+5803 &  0.061  &  25     &  275.2   &  334   &  215   &  3016  &  4362  &  1722  &  151    &  28.4 &  3887  &  4234  &  1706  &  88.4   &  0      &  3640  &  4036  &  0\\
  77 & SDSSJ1105+0745 &  0.074  &  7.83   &  253.4   &  24.6  &  46.3  &  3702  &  7001  &  2847  &  13.5   &  9.4  &  5121  &  7397  &  3150  &  2.57   &  8.4    &  2838  &  5754  &  2710\\
  78 & SDSSJ1623+4104 &  0.045  &  15.86  &  32.67   &  78.4  &  272   &  1645  &  3515  &  1388  &  56.9   &  126  &  1789  &  3801  &  1337  &  27.9   &  45.4   &  1851  &  3454  &  1571\\
  79 & SDSSJ0830+3405 &  0.07   &  21.87  &  454.94  &  75.2  &  84.6  &  4570  &  5859  &  3671  &  53.2   &  0    &  5491  &  5424  &  0     &  30.8   &  0      &  4874  &  5439  &  0\\
  80 & SDSSJ1619+4058 &  0.034  &  ...    &  ...     &  ...   &  ...   &  ...   &  ...   &  ...   &  ...    &  ...  &  ...   &  ...   &  ...   &  ...    &  ...    &  ...   &  ...   &  ...\\
  81 & SDSSJ0857+0528 &  0.038  &  2.64   &  29.51   &  27.2  &  55.5  &  1828  &  3531  &  1394  &  8.26   &  14.7 &  2344  &  3801  &  1788  &  0      &  17.2   &  2036  &  0     &  2151\\
  82 & SDSSJ1613+3717 &  0.059  &  9.27   &  227.9   &  40.1  &  55.1  &  3336  &  5935  &  2399  &  21.6   &  8.94 &  4997  &  6370  &  2468  &  4.97   &  12.7   &  2529  &  6330  &  2301\\
  83 & SDSSJ1025+5140 &  0.062  &  9.42   &  34.11   &  47.3  &  96.9  &  1873  &  4617  &  1405  &  22.6   &  27.6 &  1912  &  4575  &  1194  &  5.56   &  16.2   &  2406  &  4258  &  2339\\
  84 & SDSSJ1016+4210 &  0.055  &  14.11  &  41.73   &  94.5  &  267   &  1645  &  3576  &  1318  &  44.6   &  112  &  1851  &  3904  &  1463  &  24     &  47.6   &  1789  &  3452  &  1572\\
  85 & SDSSJ1128+1023 &  0.051  &  10.94  &  16.71   &  99.5  &  248   &  1142  &  2601  &  880   &  39.7   &  64.7 &  1357  &  2783  &  974   &  ...    &  ...    &  ...   &  ...   &  ...\\
  86 & SDSSJ1300+5641 &  0.072  &  13.98  &  54.01   &  54.5  &  83.1  &  1645  &  3151  &  1171  &  23.3   &  20.3 &  2221  &  3227  &  1494  &  ...    &  ...    &  ...   &  ...   &  ...\\
  87 & SDSSJ1538+4440 &  0.041  &  3.79   &  66.77   &  15    &  41.8  &  2787  &  5859  &  2273  &  9.7    &  15   &  3763  &  5548  &  2938  &  0      &  11.8   &  2900  &  0     &  3165\\
  88 & SDSSJ1342+5642 &  0.073  &  4.99   &  52.18   &  34.7  &  35.6  &  1736  &  3120  &  1074  &  13.9   &  8.59 &  2776  &  3390  &  2034  &  0      &  10.4   &  2036  &  0     &  2276\\
  89 & SDSSJ1344+4416 &  0.055  &  8.3    &  17.68   &  41.7  &  113   &  1188  &  2961  &  964   &  26.1   &  49.7 &  1480  &  3185  &  1070  &  7.61   &  26.2   &  1419  &  3537  &  1300\\
  90 & SDSSJ1554+3238 &  0.049  &  17.46  &  384.87  &  60.4  &  82.8  &  4387  &  5859  &  3594  &  38     &  0    &  5491  &  5397  &  0     &  0      &  19.3   &  2653  &  0     &  2836\\
\enddata
\end{deluxetable}

\begin{thebibliography}{}
\bibitem[Barger et al.(2005)]{2005AJ....129..578B} Barger, A.~J., Cowie,
L.~L., Mushotzky, R.~F., Yang, Y., Wang, W.-H., Steffen, A.~T., \&
Capak, P.\ 2005, \aj, 129, 578
\bibitem{Col06} Collin, S., et al. 2006, A\&A, 456, 75C
\bibitem[Greene
\& Ho(2006)]{2006ApJ...641L..21G} Greene, J.~E., \& Ho, L.~C.\ 2006, \apjl, 641, L21
\bibitem[Bian
\& Zhao(2004)]{2004MNRAS.352..823B} Bian, W., \& Zhao, Y.\ 2004, \mnras, 352, 823
\bibitem{Ben06} Bentz, M.C., et al. 2006, ApJ, 644, 133
\bibitem{Bin93} Binette, L., et al. 1993, ApJ, 414, 535
\bibitem{Bon06} Bon, E., Popovi\'c, L.\u C, et al. 2006, NewAR, 50, 716

\bibitem[Brotherton(1996)]{brotherton96}Brotherton, M. S. 1996, \apjs, 102, 1
\bibitem[Collin et al. (2006)]{}Collin, S., Kawaguchi, T., Peterson, B.M., \& Vestergaard, M. 2006, A\&A, 456, 75
\bibitem{Cor03} Corbett, E.A., et al. 2003, MNRAS, 343, 705
\bibitem{Dec08} Decarli, R., et al. 2008, MNRAS, 386, 15
\bibitem{Eli06} Elitzur, M. 2006, in The Central Engine of Active Galactic Nuclei, ed. L. C. Ho \&
J.-M. Wang (San Francisco: ASP), 3
\bibitem{Elv94} Elvis, M., et al., 1994, ApJS, 95, 1
\bibitem{Era03} Eracleous, M. \& Halpern, Jules P. 2003, ApJ, 599,
886
\bibitem[Ferrarese
\& Ford(2005)]{2005SSRv..116..523F} Ferrarese, L., \& Ford, H.\ 2005, Space Science
Reviews, 116, 523
\bibitem[Gebhardt et al.(2000)]{2000ApJ...539L..13G} Gebhardt, K., et al.\
2000, \apjl, 539, L13
\bibitem{Gon99} Goncalves, A.C., et al. 1999, A\&A, 341, 662
\bibitem{Gre05} Greene, J.E. \& Ho, L.C. 2005, ApJ, 630, 122
\bibitem[Grupe
\& Mathur(2004)]{2004ApJ...606L..41G} Grupe, D., \& Mathur, S.\ 2004, \apjl, 606, L41
\bibitem[Hasinger(2004)]{2004NuPhS.132...86H} Hasinger, G.\ 2004, Nuclear
Physics B Proceedings Supplements, 132, 86
\bibitem[Hasinger(2005)]{2005gbha.conf..418H} Hasinger, G.\ 2005, Growing
Black Holes: Accretion in a Cosmological Context, 418
\bibitem{hu08}Hu, C. et al. 2008, ApJ, 683, L115

\bibitem{Jol85}Joly, M. et al. 1985, A\&A, 152, 282
\bibitem{Kas00} Kaspi, S., et al. 2000, ApJ, 533, 631
\bibitem{Kas05} Kaspi, S., et al. 2005, ApJ, 629, 61K
\bibitem{Kob93} Kobayashi, Y., Sato, S., Yamashita, T., Shiba, H., Takami, H., 1993, ApJ, 404,
94
\bibitem[Komossa
\& Xu(2007)]{2007ApJ...667L..33K} Komossa, S., \& Xu, D.\ 2007, \apjl, 667, L33
\bibitem{Kom08} Komossa, S., et al. 2008, ApJ, 680, 926
\bibitem[Komossa(2008)]{2008RMxAC..32...86K} Komossa, S.\ 2008, Revista Mexicana de Astronomia y Astrofisica Conference Series, 32, 86
\bibitem{Kro78} Krolik, J.H., et al. 1978, ApJS, 37, 459
\bibitem{Kro88} Krolik, J. H. \& Begelman, M. C. 1988, ApJ, 329, 702
\bibitem{Mur07} La Mura, G., et al. 2007, ApJ, 671, 104
\bibitem[Laor(2004)]{2004ASPC..311..169L} Laor, A.\ 2004, AGN Physics with
the Sloan Digital Sky Survey, ASP Conference Series, 311, 169
\bibitem{Law91} Lawrence, A. 1991, MNRAS, 252, 586
\bibitem[Magorrian et al.(1998)]{1998AJ....115.2285M} Magorrian, J., et al.\ 1998, \aj, 115, 2285
\bibitem{Mai07} Maiolino, R., et al. 2007, A\&A, 468, 979
\bibitem{Mas96} Mason, K.O., et al. 1996, MNRAS, 283, L26
\bibitem{Mar09} Marziani, P., et al. 2009, A\&A, 495, 83M
\bibitem[Marconi
\& Hunt(2003)]{2003ApJ...589L..21M} Marconi, A., \& Hunt, L.~K.\
2003, \apjl, 589, L21
\bibitem{Mul08} Mullaney, J.R. \& Ward, M.J. 2008, MNRAS, 385, 53
\bibitem{Ne208} Nenkova, M. et al. 2008a, ApJ, 685, 147
 (2008arXiv:0806.0511)
\bibitem{Net93} Netzer, H. \& Laor, A. 1993, ApJ, 404, L51
\bibitem[Onken et al. (2004)]{}Onken, C.A., et al. 2004, ApJ, 615, 645
\bibitem{Ost89} Osterbrock, D.E. Astrophysics of Gaseous Nebulae and Active Nuclei, 1989 University Science Books press, P 337-338.
\bibitem{Pet98} Peterson, B.M., et al. 1998, ApJ, 501, 82
\bibitem{Pet99} Peterson, B.M., et al. 1999, ASPC, 175, 41
\bibitem[Peterson \& Wandel(1999)]{peterson99}Peterson, B. M., \& Wandel, A. 1999, \apj, 521, L95
\bibitem{Pet04} Peterson, B.M., et al. 2004, ApJ, 613, 682
\bibitem{Pet06} Peterson, B.M. 2006, Memorie della Societa Astronomica
 Italiana, 77, 581

\bibitem[Peterson (2007)]{}Peterson, B. M. 2007, in The Central Engine of Active Galactic Nuclei, ed. L. C. Ho \&
J.-M. Wang (San Francisco: ASP), 3
\bibitem{Pop08} Popovi\'c, L.\u C, et al. 2008, RmxAC, 32, 99
\bibitem{Sim08} Simpson, C. 2005, MNRAS, 360, 565
\bibitem{slu07} Sluse, D., et al. 2007, A\&A, 468, 885
\bibitem{slu08} Sluse, D., et al. 2008, RMxAC, 32, 83
\bibitem{Str03} Strateva, Iskra V. et al. 2003, AJ, 126, 1720
\bibitem[Steffen et al.(2003)]{2003ApJ...596L..23S} Steffen, A.~T., Barger,
A.~J., Cowie, L.~L., Mushotzky, R.~F., \& Yang, Y.\ 2003, \apjl, 596, L23
\bibitem[Steffen et al.(2004)]{2004AJ....128.1483S} Steffen, A.~T., Barger,
A.~J., Capak, P., Cowie, L.~L., Mushotzky, R.~F., \& Yang, Y.\ 2004, \aj, 128, 1483
\bibitem{Sug06} Suganuma, M., et al. 2006, ApJ, 639, 46
\bibitem[Sulentic et al.(2000b)]{sulentic00b}Sulentic, J. W., Marziani, P., Zwitter, T., Dultzin-Hacyan, D., \& Calvani, M. 2000, \apj, 545, L15
\bibitem{She08} Shen, J.J., et al. 2008, AJ, 135, 928
\bibitem{Tre02} Tremaine, S., et al. 2002, ApJ, 574, 740
\bibitem{Ued03} Ueda, Y., Akiyama, M., Ohta, K., \& Miyaji, T. 2003, ApJ, 598, 886
\bibitem{Urr95} Urry, C.M., \& Padovani, P. 1995. PASP, 107, 803
\bibitem{Ves06} Vestergaard, M. \& Peterson, B.M. 2006, ApJ, 641, 689
\bibitem{Ver02} V\'eron,P ., et al. 2002, A\&A, 384, 826
\bibitem{Wan05} Wang, J-M., et al. 2005, ApJ, 627, 5

\bibitem[Wang
\& Zhang(2007)]{2007ApJ...660.1072W} Wang, J.-M., \& Zhang, E.-P.\ 2007, \apj, 660,
1072
\bibitem[Wang
\& Lu(2001)]{2001A&A...377...52W} Wang, T., \& Lu, Y.\ 2001, \aap, 377, 52
\bibitem[Williams et al. (2004)]{}Williams, R. J., Mathur, S., \& Pogge, R. W. 2004, ApJ, 610, 737
\bibitem[Zhou et al. (2006)]{2006ApJS..166..128Z} Zhou, H., Wang, T., Yuan,
W., Lu, H., Dong, X., Wang, J., \& Lu, Y.\ 2006, \apjs, 166, 128
\bibitem[Zhang et al. (2009)]{} Zhang, Kai., Wang, Tinggui., et al.
2009, arXiv:0902.4390.

\bibitem[zhangwm] Zhang, W. M., Soria, R. Zhang, S. N. Swartz, D. A., Liu,
J. 2009, accepted for publication in ApJ (arXiv:0904.1091)
\bibitem[Zhu
\& Zhang(2009)]{} Zhu, L., \& Zhang, S.N. 2009, Science in China (G), in press (arXiv0907.1942Z)
% \bibitem{}
% \bibitem{}

\end{thebibliography}
\end{document}